\documentclass[12pt]{article}

\usepackage{amsmath,amssymb,cite,bm}
\usepackage{graphicx}

\makeatletter
 
  \@addtoreset{equation}{section}
 \makeatother

\newcommand{\n}{\nonumber \\}
\newcommand{\Tr}{\mathrm{Tr}}

\newcommand{\be}{\begin{equation}}
\newcommand{\ee}{\end{equation}}
\newcommand{\bea}{\begin{eqnarray}}
\newcommand{\eea}{\end{eqnarray}}
\newcommand{\beann}{\begin{eqnarray*}}
\newcommand{\eeann}{\end{eqnarray*}}

\newcommand{\ba}{\begin{array}}
\newcommand{\ea}{\end{array}}

\allowdisplaybreaks

\begin{document}

\setlength{\oddsidemargin}{0cm}
\setlength{\baselineskip}{7mm}

\begin{titlepage}
\renewcommand{\thefootnote}{\fnsymbol{footnote}}
\begin{normalsize}
\begin{flushright}
\begin{tabular}{l}
DIAS-STP-17-13, UTHEP-709, YITP-17-123
\end{tabular}
\end{flushright}
  \end{normalsize}

~~\\

\vspace*{0cm}
    \begin{Large}
%    \begin{bf}
       \begin{center}
         {Spherical transverse M5-branes from the plane wave matrix model
% Spherical fivebranes?
% Fivebranes?
}
       \end{center}
%    \end{bf}   
    \end{Large}
\vspace{0.5cm}

\begin{center}
Yuhma A{\sc sano}$^{1)}$\footnote
            {
e-mail address : 
yuhma@stp.dias.ie},
Goro I{\sc shiki}$^{2),3)}$\footnote
            {
e-mail address : 
ishiki@het.ph.tsukuba.ac.jp},
Shinji S{\sc himasaki}$^{4)}$\footnote
            {
e-mail address : 
shinji.shimasaki@keio.jp} and 
Seiji T{\sc erashima}$^{5)}$\footnote
            {
e-mail address : 
terasima@yukawa.kyoto-u.ac.jp}

\vspace{0.5cm}
\small{
     $^{ 1)}$ {\it School of Theoretical Physics, Dublin Institute for Advanced Studies, }\\
               {\it 10 Burlington Road, Dublin 4, Ireland}\\

     $^{ 2)}$ {\it Tomonaga Center for the History of the Universe,\\ University of Tsukuba, }
               {\it Tsukuba, Ibaraki 305-8571, Japan}\\
                   
     $^{ 3)}$ {\it Graduate School of Pure and Applied Sciences, University of Tsukuba, }\\
               {\it Tsukuba, Ibaraki 305-8571, Japan}\\

     $^{ 4)}$ {\it Research and Education Center for Natural Sciences, Keio University, }\\
               {\it Hiyoshi 4-1-1, Yokohama, Kanagawa 223-8521, Japan}\\

     $^{ 5)}$ {\it Yukawa Institute for Theoretical Physics, Kyoto 
University, }\\
               {\it Kyoto 606-8502, Japan}\\
}

\end{center}

\vspace{0.3cm}

\begin{abstract}
\noindent
%The plane wave matrix model (PWMM) is conjectured to give a formulation of 
%M-theory on the pp-wave background. In the M-theory, there exist spherical 
%M5-branes with vanishing light cone energy. These fivebranes have been 
%conjectured to be described as some vacuum states in PWMM. In this paper, 
%we test this conjecture by directly analyzing the strong coupling regime of 
%PWMM based on the localization method. We find that the eigenvalue 
%distribution of the SO(6) scalar fields forms a five-dimensional spherical 
%shell in an appropriate limit. Furthermore, we show that the radius of the 
%distribution exactly agrees with the radius of the M5-brane in the M-theory.
We consider matrix theoretical description of transverse M5-branes in M-theory on the 11-dimensional maximally supersymmetric pp-wave background.
We apply the localization to the plane wave matrix model (PWMM) 
and show that the transverse spherical fivebranes with zero 
light cone energy in M-theory are realized as the distribution 
of low energy moduli of the $SO(6)$ scalar fields in PWMM.
\end{abstract}
\vfill

\end{titlepage}
\vfil\eject

\setcounter{footnote}{0}

%\tableofcontents

\section{Introduction}

Matrix models are conjectured to give nonperturbative formulations 
of M-theory \cite{Banks:1996vh}. 
This formulation is expected to realize a second 
quantization of M-theory, which contains 
all the fundamental objects in the theory. 
However, the description of states with M5-branes in the matrix models has 
not been established yet. 
Understanding this problem will shed light on the 
matrix-model formulation of M-theory.

In this paper, we focus on M-theory defined on the maximally supersymmetric 
pp-wave solution of the 11-dimensional supergravity 
and consider the description of certain M5-branes living in this geometry in 
terms of the matrix model.
On this background, there exist stable spherical 
M2- and M5- branes with zero light cone energy.
According to the matrix-model conjecture, 
objects with zero light cone energy should be 
realized as vacuum states in the corresponding matrix model.
Hence, these spherical branes should also be realized 
as certain vacuum states in the matrix model.
In this paper, we investigate this relation in detail 
by using the localization method. 

The matrix model for M-theory on the pp-wave background 
is called the plane wave matrix model (PWMM) \cite{Berenstein:2002jq}.
This model is given by a mass deformation of the BFSS matrix model
\cite{Banks:1996vh}, 
where the mass parameter is proportional to the three form flux on
the pp-wave geometry.
Because of the mass deformation, PWMM possesses many discretely 
degenerate vacua, unlike the BFSS matrix model. 
The relation between these vacua and objects 
with vanishing light cone energy in M-theory was proposed in 
\cite{Berenstein:2002jq, Maldacena:2002rb}.
Here, in particular, the vacua corresponding to the 
above mentioned spherical M5-brane and its multiple generalization 
were also specified.
For the case of a single M5-brane,
this correspondence was tested by comparing the BPS protected mass spectra of 
PWMM with that of the M5-brane \cite{Maldacena:2002rb}.

Let us review this proposal in more detail.
The vacua of PWMM, which preserve all the supersymmetry, 
are given by the fuzzy sphere \cite{Madore:1991bw} and are labeled by 
$N$-dimensional representations of the $SU(2)$ Lie algebra, where
$N$ is the matrix size of PWMM.
Generally, the classical vacuum configuration in PWMM takes the form of
\begin{align}
X_i \propto L_i, \;\; (i=1,2,3)
\label{vacuum of PWMM}
\end{align}
where $X_i$ are the $SO(3)$ scalar fields in PWMM and the other fields 
are vanishing at the vacuum.
$L_i$ are $N$-dimensional 
representation matrices of the $SU(2)$ generators. Any $N$-dimensional 
representation gives a supersymmetric vacuum and,
in general, the representation is reducible. Then, one 
can make an irreducible decomposition:
\begin{align}
L_i = \bigoplus_{s=1}^\Lambda L_i^{[n_s]} \otimes 1_{N_2^{(s)}}.
\label{irrdec}
\end{align}
Here, $L_i^{[n_s]}$ are the generators in the $n_s$-dimensional 
irreducible representation and 
$N_2^{(s)}$ represents the multiplicity of the $s$th representation.
Hence, the vacua can be labeled by a set of integers
$\{ \Lambda, N_2^{(s)}, n_s | s=1, 2, \cdots, \Lambda \}$ satisfying
$\sum_{s=1}^{\Lambda}n_s N_2^{(s)}=N$.

From this structure of the vacua, we can immediately find the structure 
of the spherical M2-brane in M-theory. 
The fuzzy sphere is a regularization of a 
smooth two-dimensional sphere. In the commutative limit, where
$N_2^{(s)}$ are fixed while $n_s$ go to infinity, smooth 
two-spheres are realized from the fuzzy sphere. 
One can naturally expect that this smooth sphere is the 
spherical M2-brane with zero light cone energy.

On the other hand, in \cite{Berenstein:2002jq}, 
the spherical M5-brane was conjectured to be realized as the 
trivial vacuum of PWMM, where all the fields are vanishing. 
This is the case where the representation 
in (\ref{vacuum of PWMM}) is a direct sum of $N$ 
trivial representations.
Furthermore, the conjecture was generalized to the case of 
multiple spherical M5-branes \cite{Maldacena:2002rb}. 
In these conjectures, the M5-branes 
are considered to be realized in the limit such that
$n_s$ are fixed and  $N_2^{(s)}$ go to infinity in
(\ref{irrdec}).

\begin{figure}[t]
\includegraphics[width=15cm,clip]{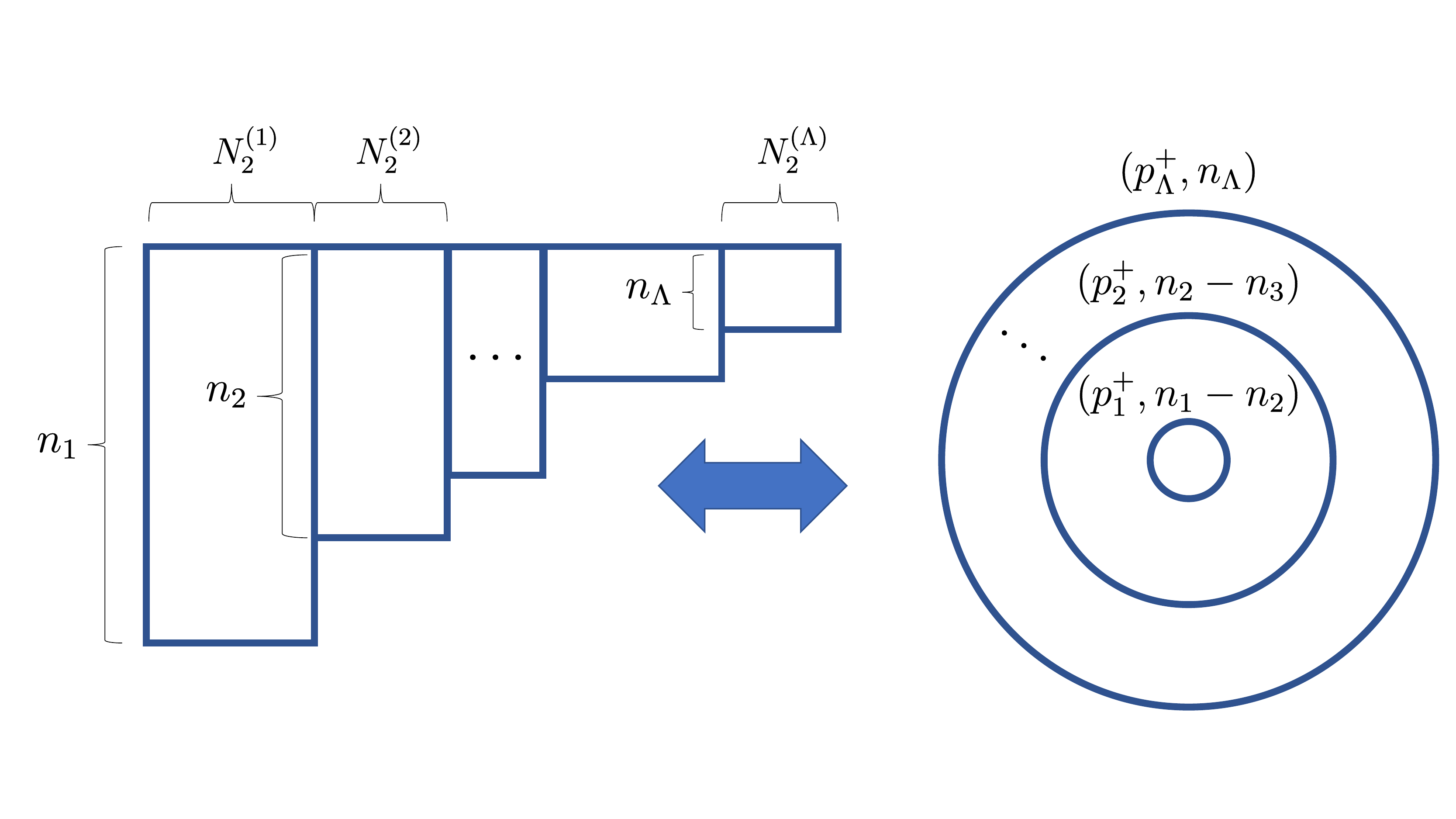}
\label{young diagram}
\caption{Correspondence between partitions and configurations of 
M5-branes.}
\end{figure}

In order to describe this limit more precisely, let us introduce 
Young diagrams associated with the partition of (\ref{irrdec}).
In the decomposition (\ref{irrdec}), we assume that 
$n_1> n_2 > \cdots > n_\Lambda$ without loss of generality.
Then we consider a Young diagram which consists of $N_2^{(1)}$ 
columns with length $n_1$, $N_2^{(2)}$ 
columns with length $n_2$, and so on. See Fig.~\ref{young diagram}.
The conjecture states that when the lengths of some rows go to infinity,
such rows correspond to the spherical M5-branes, where the light cone momentum
of each M5-brane is proportional to the length of each row.
For example, in Fig.~\ref{young diagram}, 
let us consider the limit where all $N_2^{(s)}$ go to infinity with the same order while all $n_s$ are fixed. 
This limit corresponds to a situation in M-theory such that 
there are $\Lambda$ stacks of spherical M5-branes, where
the $s$th stack is made of $n_s-n_{s+1}$ M5-branes\footnote{
For $s=\Lambda$, we define $n_{\Lambda+1}:=0$.} with light cone momentum 
\begin{align}
p_s^+=\sum_{r=1}^s N_2^{(r)}/R,
\label{pplus for sth stack}
\end{align}
where $R$ is the radius of the 
light like circle\footnote{As we will see in the next section, the radius of a single 
(i.e. not coincident) M5-brane is proportional to $(p^+)^{1/4}$. 
Thus, larger $p^+$ gives a larger radius. 
Though this relation had never been derived for coincident M5-branes,
our results discussed below shows that this is also true for coincident 
M5-branes. Fig.~\ref{young diagram} is 
based on this picture, so that the $s$th 
stack has a larger radius than $(s-1)$th stack.}.
Note that the total light cone momentum is given by 
$p^+ = \sum_{s=1}^{\Lambda}(n_s-n_{s+1})p_s^{+}$ and this is equal to 
$N/R$.
Note also that 
$N_5:={\rm max}\{n_s | s=1,2, \dots \}=n_1$ 
corresponds
to the total number of M5-branes.

This conjecture is highly nontrivial. For example, 
let us consider the simplest partition with $\Lambda=1, n_1=1, N_2^{(1)}=N$,
which corresponds to the trivial vacuum of PWMM.
At the classical level, the 
vacuum configuration is just vanishing, 
so that we can not see any structure of the M5-brane.
For example, it looks seemingly impossible to 
reproduce geometric information of the spherical M5-brane
(the radius etc.) from the trivial configuration.
Nevertheless, the conjecture claims that a single spherical M5-brane 
is realized in the trivial vacuum.

To bridge this gap, 
one needs to recall that M-theory is conjectured to be 
realized in an appropriate large-$N$ limit of PWMM, where 
the coupling constant also becomes very large as 
the matrix size $N$ goes to infinity.
%and
%the light-cone momentum $p^+$ in M-theory is held fixed. 
%In terms of the gauge coupling constant $g$ in PWMM, 
%this is translated to the limit where $N \rightarrow \infty $ with 
%$g^2/N^3$ is fixed.  
Thus, one has to deal with the strongly coupled regime of PWMM, in 
order to understand the description of M5-branes.
In the strong coupling region, there 
must be a large quantum fluctuation around the classical vacuum 
configuration. Thus, typical configurations of matrices will be
very different from the classical configuration. 
There is a possibility that the spherical M5-branes are formed 
as a typical configuration of matrices in the strong coupling 
region of PWMM\footnote{
See also \cite{Lozano:2005kf} for the description of M5-branes in a 
different matrix model.}.

In this paper, we investigate this possibility by directly studying the 
strong coupling regime of PWMM.
The limit we consider is 
\begin{align}
N_2^{(s)} \rightarrow \infty,  \;\; n_s\; {\rm fixed},  
\;\; N_2^{(s)}/N_2^{(t)} \; {\rm fixed} 
\label{limit M5}
\end{align}
for any $s,t=1,2,\cdots,\Lambda$. This limit corresponds to
$\Lambda$ stacks of M5-branes with different radii as shown in 
Fig.~\ref{young diagram}.
In addition, we also scale the coupling constant of PWMM in such a way that 
the M5-branes decouple with the bulk gravity and 
only the degrees of freedom on the M5-branes become relevant 
\cite{Maldacena:2002rb}. 
This decoupling limit turns out to be
the strong coupling limit in the 't Hooft limit of PWMM, as we will describe 
in the next section.
In this decoupling limit, we apply the localization to PWMM and 
reduce some BPS correlation functions to certain eigenvalue integrals. 
By evaluating the eigenvalue integral, 
we argue that the eigenvalue distribution of the low energy modes of 
the $SO(6)$ scalar fields forms 
$\Lambda$ stacks of spherical shells and coincides with the expected 
configuration of 
the spherical M5-branes in M-theory\footnote{A part of this result
was briefly reported in the letter \cite{Asano:2017xiy} for the case of concentric M5-branes. In this paper, we not only describe the technical details of 
\cite{Asano:2017xiy} but also generalize the result of 
\cite{Asano:2017xiy} to the most general configurations of 
the spherical M5-branes.}.
In particular, we show that, for a single M5-brane,
the radius of the shell completely agrees with the value 
computed by using the classical Dirac-Nambu-Goto action of a single M5-brane.
This result strongly supports the proposal 
of \cite{Maldacena:2002rb} and shows that PWMM indeed contains the 
multiple M5-brane states.
We also apply the same argument to M2-branes and show that 
the spherical M2-brane can be described in a similar way using the 
eigenvalue integral.

%In this paper, we investigate this possibility focusing on 
%a simple case of $\Lambda=1$, namely, the case with
%\begin{align}
%L_i = L_i^{[N_5]} \otimes 1_{N_2}.
%\label{specific vacuum}
%\end{align}
%Here, the total matrix size is given by $N=N_2N_5$. 
%According to the above correspondence, this vacuum corresponds to the 
%$N_5$ M5-branes, each of which carries the light cone momentum $N_2$.

%The scaling limits of our interest is 
%\begin{align}
%&{\mbox{M2-brane limit}}: \;\;N_5 \rightarrow \infty,  \;\; N_2\; {\rm fixed},
%\nonumber\\
%&{\mbox{M5-brane limit}}: \;\;N_2 \rightarrow \infty,  \;\; N_5\; {\rm fixed}.
%\label{M2-M5-limit}
%\end{align}
%In addition, for each limit (\ref{M2-M5-limit}), 
%we also scale the coupling constant of PWMM in such a way that 
%only the degrees of freedom on these branes become relevant 
%\cite{Maldacena:2002rb}.
%%In these decoupling limits, the VEV of the square of the scalar fields 
%%in PWMM should correspond to the radius of the corresponding spherical 
%%branes.
%The decoupling limit for the M5-branes turns out to be
%the strong coupling limit in the 't Hooft limit of PWMM, as we will describe 
%in the next section.
%%In particular, the case of $N_5=1$ is just the strong coupling limit of 
%%the trivial vacuum, and this corresponds to the single M5-brane.
%%According to the above conjecture, this case should describe 
%%the single spherical M5-brane.

This paper is organized as follows. 
In section \ref{M-theory on the pp-wave background},
we review M-theory on the pp-wave background. 
We show that there exist spherical M2- and M5- branes 
with zero light cone energy on this background.
We also compute the radii of these objects.
In section \ref{The plane wave matrix model},
we review PWMM. 
In section \ref{Spherical M5-branes from PWMM},
we apply the localization to PWMM and evaluate the 
moduli distribution of scalar fields.
We show that the distribution agrees with the configuration of the 
spherical M5-branes. 
In section
\ref{Spherical M2-branes from PWMM}, we consider the case of M2-branes. 
In section \ref{Summary},
we summarize our results and discuss the low energy theory of PWMM.
%which was proposed as a model describing 
%a certain BPS sector of ${\cal N}=4$ SYM in the strong coupling regime.

\section{M-theory on the pp-wave background}
\label{M-theory on the pp-wave background}
In this section, we review M-theory on the 
maximally supersymmetric plane wave background in the 11-dimensional 
supergravity. The background geometry is given by
\begin{align}
ds^2 &=g_{\mu \nu}dx^\mu dx^\nu 
= -2dx^+ dx^- + \sum_{A=1}^9 dx^A dx^A -
\left( 
\frac{\mu^2 }{9}\sum_{i=1}^3x^ix^i +
\frac{\mu^2 }{36}\sum_{a=4}^9x^ax^a
\right) dx^+dx^+,
\nonumber\\
F_{123+} &= \mu,
\label{pp bg}
\end{align}
where $\mu$ is the flux parameter of the three form field\footnote{
Throughout this paper, we mainly use the notation that 
$\mu, \nu =0,1,2,\cdots,10$, 
$A,B=1,2,\cdots,9$, $i,j=1,2,3$ and $a,b=4,5,\cdots,9$.
}. 
We will see that spherical M2-brane and M5-brane exist as the lowest energy 
states with respect to the light cone Hamiltonian.
We refer the method in \cite{Taylor:2001vb} 
for the calculation in this section.

\subsection{Spherical M2-brane}
\label{spherical M2-brane}
We first consider a single M2-brane in the background (\ref{pp bg}).
The bosonic part of the M2-brane action is given by 
the Dirac-Nambu-Goto action plus a Chern-Simons term as
\begin{align}
S_{\rm M2}= -T_{\rm M2} \int d^3 \sigma \sqrt{-{\rm det}h_{\alpha \beta}}
+ T_{\rm M2} \int C_3.
\label{M2 action}
\end{align}
Here, $h_{\alpha \beta}$ 
is the induced metric, 
\begin{align}
h_{\alpha \beta}=g_{\mu\nu }(X)
\partial_{\alpha }X^\mu
\partial_{\beta }X^\nu,
\end{align}
for the embedding function $X^\mu (\sigma) $.
The overall constant $T_{\rm M2}$ in (\ref{M2 action}) 
is the tension of M2-brane given by
\begin{align}
T_{\rm M2}= \frac{1}{(2\pi)^2 l_p^3},
\label{M2-tension}
\end{align}
where $l_p$ stands for the Planck length.
By introducing a symmetric auxiliary field $\gamma_{\alpha \beta}$,
we rewrite the action into the Polyakov type:
\begin{align}
S_{\rm M2}=-\frac{T_{M2}}{2}
\int d^3 \sigma 
\sqrt{-\gamma}
\left(
\gamma^{\alpha \beta}
g_{\mu \nu}(X)\partial_{\alpha }X^\mu
\partial_{\beta }X^\nu
-1 \right)
+
T_{\rm M2} \int C_3.
\end{align}

This action has a diffeomorphism symmetry for the worldvolume coordinates 
$\sigma^\alpha= (\sigma^0, \sigma^1, \sigma^2)$ of the membrane. 
If we consider an M2-brane with topology $R\times \Sigma$, 
where $R$ is the time direction and $\Sigma$ is a Riemann surface,
we can fix this symmetry by putting
\begin{align}
\gamma_{0a}=0, \;\;\; \gamma_{00}= -\frac{4}{\nu^2}{\rm det}h_{ab},
\label{gauge fixing condition}
\end{align}
where $a,b=1,2$ and the determinant is taken in this $2\times 2$
subspace.
$\nu$ is a constant which will be related to the light cone momentum 
of the M2-brane below.
Then, the action becomes
\begin{align}
S_{\rm M2}&= \frac{T_{\rm M2}\nu }{4}
\int d^3 \sigma \left(
h_{00}-\frac{4}{\nu^2}{\rm det}h_{ab}
\right)+
T_{\rm M2} \int C_3
\nonumber\\
& =\frac{T_{\rm M2}\nu }{4}
\int d^3 \sigma \left(
-2\dot{X}^- +(\dot{X}^A)^2
-\frac{\mu^2}{9}(X^i)^2
-\frac{\mu^2}{36}(X^a)^2
-\frac{2}{\nu^2}\{X^A,X^B \}^2
\right)+
T_{\rm M2} \int C_3.
\label{M2 action after gauge fixing}
\end{align}
Here, in the second line, we have introduced a 
canonical Poisson bracket on the membrane defined 
by $\{f, g\}= \epsilon^{ab} (\partial_a f)(\partial_b g) $ for 
each fixed $\sigma^0$.
In terms of the Poisson bracket, 
the Chern-Simons term can be written as
\begin{align}
\int C_3 = \frac{\mu}{6}
\int d^3 \sigma \epsilon_{ijk}
X^i \{ X^j, X^k \}.
\end{align}
The gauge fixing condition (\ref{gauge fixing condition})
as well as the equation of motion of the auxiliary field 
produce the following constraints:
\begin{align}
&g_{\mu \nu}\dot{X}^\mu \dot{X}^\nu 
=-\frac{2}{\nu^2}g_{\mu \nu}g_{\rho \sigma}
\{X^\mu, X^\rho \}\{X^\nu, X^\sigma \}
\nonumber\\
&g_{\mu \nu}\dot{X^\mu}\partial_a X^{\nu} =0.
\label{constraint 1}
\end{align}
From the second constraint, it also follows that
\begin{align}
\{ g_{\mu \nu} \dot{X}^\mu, X^\nu \} =0.
\label{constraint 2}
\end{align}
Thus, the system is reduced to the theory 
(\ref{M2 action after gauge fixing})
with these constraints imposed.

The constraints (\ref{constraint 1}) can be 
explicitly solved in the light cone gauge, 
%where we choose the world volume coordinate $\sigma^0$ as
\begin{align}
X^+(\sigma )= \sigma^0.
\end{align}
Here, we have defined $X^\pm $ by 
\begin{align}
X^\pm = \frac{1}{\sqrt{2}}
(X^0 \pm X^{10}).
\end{align}

We then consider the Hamilton formalism. 
We denote by $P^\mu $ the canonical 
conjugate momentum of $X^\mu $.
The total light cone momentum is then given by
\begin{align}
p^+ = \int d^2 \sigma P^+ =
2\pi \nu T_{\rm M2},
\end{align}
where we have chosen the spacial coordinates such that 
they have a volume $\int d^2 \sigma = 4\pi $.
This relates the constant ${\nu}$ to the light cone momentum.
The Hamiltonian is given by 
\begin{align}
H_{\rm M2}
= &\int d^2\sigma
\left[
\frac{V_2}{2p^+}
\left( P_A^2 +
\frac{T_{\rm M2}^2}{2}
\{X^A, X^B\}^2
\right) 
\right.
\nonumber\\
&\left.
+\frac{p^+}{2V_2}
\left(
\frac{\mu^2}{9}(X^i)^2
+
\frac{\mu^2}{36}(X^a)^2
\right)
-\frac{\mu T_{\rm M2}}{6}
\epsilon_{ijk}X^i \{X^j, X^k \}
\right],
\label{HamM2}
\end{align}
where, $V_2$ is the volume of the unit sphere, $V_2 = 4\pi$.
The remaining constraint (\ref{constraint 2})
is written in terms of the transverse components $X^A$
as 
\begin{align}
\{P^A, X_A\}=0.
\label{constraint after gauge fixing}
\end{align}

Now, let us consider a vacuum configuration, which 
minimizes the Hamiltonian (\ref{HamM2}).
Note that the potential for $X^i$ forms a perfect square, 
\begin{align}
\frac{p^+ \mu^2}{18 V_2}
\left(
X_i -\frac{3V_2T_{\rm M2}}{2\mu p^+}\epsilon_{ijk}
\{X^i,X^j\}\right)^2.
\label{perfect square M2}
\end{align}
From this, we find that the
vacuum configuration is given by
\begin{align}
X^i = r_{\rm M2}x^i,  \;\;\;\;\;  X^a=0,
\;\;\;\;\;  P_A=0,
\label{spherical M2 config}
\end{align}
where $x^i$ are the embedding function of the unit sphere in 
$R^3$ satisfying
\begin{align}
x^ix^i=1, \;\;\; \{x^i,x^j\}=\epsilon^{ijk}x_k.
\label{embedding of s2}
\end{align}
The radius is also determined as
\begin{align}
r_{\rm M2}= \frac{\mu p^+}{12 \pi T_{\rm M2}}.
\label{radius M2}
\end{align}
The configuration (\ref{spherical M2 config}) obviously has  
the spherical shape.
Thus we see that, in M-theory on the pp-wave background, 
there exists a spherical zero energy M2-brane with the radius
given by (\ref{radius M2}).

\subsection{Spherical M5-brane}
Then, let us consider a single M5-brane. 
We start from the bosonic part of the action,
\begin{align}
S_{\rm M5}= -T_{\rm M5} \int d^6 \sigma \sqrt{-{\rm det}h_{\alpha \beta}}
+ T_{\rm M5} \int C_6,
\label{action M5}
\end{align}
where $dC_6 = *F_4$ and the tension is written as
\begin{align}
T_{\rm M5} = \frac{1}{(2\pi)^5 l_p^6}.
\label{M5-tension}
\end{align}
We can apply the computation in the previous subsection
to (\ref{action M5}). 
Then, we can obtain the light-cone Hamiltonian for 
the M5-brane, 
\begin{align}
H_{\rm M5}
= & \int d^5 \sigma
\left[
\frac{V_5}{2p^+}
\left( P_A^2 +
\frac{T_{\rm M5}^2}{5!}
\{X^{A_1}, \cdots, X^{A_5} \}^2
\right)
\right.
\nonumber\\
& \left. + 
\frac{p^+}{2V_5}
\left(
\frac{\mu^2}{9}(X^i)^2
+
\frac{\mu^2}{36}(X^a)^2
\right)
-\frac{\mu T_{\rm M5}}{6!}
\epsilon_{a_1a_2\cdots a_6}X^{a_1} 
\{X^{a_2},\cdots, X^{a_6} \}
\right].
\label{lc ham for M5}
\end{align}
Here, $V_5$ is the volume of the unit 5-dimensional sphere, $V_5 = \pi^3$.
The curly bracket with five entries in (\ref{lc ham for M5}) is the 
5-dimensional analogue of the Poisson bracket defined by
\begin{align}
\{f_1 ,\cdots, f_5 \} = \epsilon^{a_1\cdots a_5}
(\partial_{a_1} f_1)\cdots (\partial_{a_5}f_5).
\end{align}

We notice that the potential terms of $X^a$ forms a perfect square,
\begin{align}
\frac{p^+\mu^2}{72 V_5}
\left(
X_{a_1} -\frac{6V_5T_{\rm M5}}{5!\mu p^+}
\epsilon_{a_1a_2 \cdots a_6}
\{X^{a_2},\cdots, X^{a_6} \}\right)^2.
\end{align}
Thus, we find that the vacuum configuration is given by 
a spherical fivebrane of the form,
\begin{align}
X^i=0, \;\;\;\;\; X^{a}= r_{\rm M5}x^{a}, 
\;\;\;\;\;  P_A=0,
\end{align}
where $x^a$ are the embedding function of the unit 5-sphere into $R^6$ 
satisfying
\begin{align}
x^ax^a=1, \;\;\; \{x^{a_1},\cdots, x^{a_5}\}=
\epsilon^{a_1a_2 \cdots a_6}x_{a_6}.
\end{align}
The radius of the fivebrane is determined as
\begin{align}
r_{\rm M5}= \left( \frac{\mu p^+}{6\pi^3 T_{\rm M5}} \right)^{1/4}.
\label{radius of M5}
\end{align}

\subsection{Decoupling limits}
\label{decoupling limits}
In this paper, we focus on the limits in 
which the radii of the spherical M2- and M5- branes 
become very large and only the degrees of 
freedom on these branes survive for low energy physics\cite{Maldacena:2002rb}. 

Let us introduce the radius $r=\sqrt{x^ix^i}$ of the two sphere on which 
the M2-brane is wrapping. The metric (\ref{pp bg}) is written as
\begin{align}
ds^2 &= -2 dx^+ dx^-  -\frac{\mu^2 r^2}{9} dx^+dx^+ + r^2 d\Omega_2^2 +\cdots
\nonumber\\
&= -\frac{\mu^2 r^2 }{9}d\tilde{x}^+ d\tilde{x}^+ 
+\frac{9}{\mu^2 r^2 }d\tilde{x}^- d\tilde{x}^- + r^2 d\Omega_2^2 + \cdots,
\label{rescaled metric}
\end{align}
where $\cdots$ represents the other terms which are irrelevant in this 
discussion and we have defined $\tilde{x}^{\pm}$ by $\tilde{x}^+ = x^+ +\frac{9}{\mu^2 r^2}x^-, \tilde{x}^- = x^-$.
Note that $\tilde{x}^{\pm}$ have the periodicity 
\begin{align}
(\tilde{x}^+, \tilde{x}^-) \sim (\tilde{x}^+, \tilde{x}^-)
+({9R}/{(\mu r)^2}, R),
\end{align}
where $R$ is the radius of the original compactified circle 
along the light-like direction.
Since the shift of $\tilde{x}^+$ is much smaller than that of 
$\tilde{x}^-$ in the large-$r$ limit, this can be effectively 
regarded as a spatial compactification near the large M2-brane\footnote{One 
can also take another coordinate $(\hat{x}^+, \hat{x}^-)$ such that 
the metric becomes canonical Minkowski metric. In this coordinate, the 
both shifts of $\hat{x}^{\pm}$ are given by $R/(\mu r)$ and this looks like 
a light cone compactification. However, note 
that from $\frac{d}{d \hat{x}^+} \sim \frac{1}{\mu r} \frac{d}{dx^+}$, 
we see that the energy along $\hat{x}^+$ direction is given by 
$\frac{1}{\mu r}H$. Similarly, the momentum along $\hat{x}^-$ direction is
$\mu r p^+ +\frac{1}{\mu r}H$. In the limit discussed below, both $r $ and 
$p^+$ becomes large, so that the energy is much smaller than the spatial 
momentum. Thus, after all, 
this can be indeed regarded as a spatial compactification.
}. 
From the structure of the 
metric (\ref{rescaled metric}), 
we find that the physical radius of the M-circle is given by 
$\tilde{R} \sim R/(\mu r)$. 
In the perspective of the type IIA superstring theory, 
the spherical M2-brane wrapping on the two-sphere in 
(\ref{rescaled metric}) corresponds to a D2-brane.
The gauge coupling constant on D2-branes is given by 
$g_{YM}^2 \sim g_s l_s^{-1} $, where $g_s$ and $l_s$ are the string coupling 
and the string length. By translating this into the M-theory parameters 
using the standard dictionary, $g_s \sim (\tilde{R} /l_p)^{3/2}$ and 
$l_s \sim (l_p^3/ \tilde{R} )^{1/2}$, one can express the coupling constant as 
$g_{YM}^2 \sim \tilde{R}^2 /l_p^3 $. In the limit where the radius of the 
D-branes becomes large, it is convenient to rescale the metric, so that 
the parameter which controls the theory on D2-brane is given by the 
dimensionless coupling constant $g_{YM}^2 r_{\rm M2}$, 
where $r_{\rm M2}$ is the radius of the M2-brane.
By using (\ref{radius M2}),
The coupling constant can be expressed as 
\begin{align}
g_{YM}^2 r_{\rm M2} \sim \frac{R^2}{r_{\rm M2}\mu^2 l_p^3}.
\label{effective coupling on D2}
\end{align}
We are interested in the case where $l_p$ and $\mu$ are fixed.
Moreover, in order to have an interacting theory 
on the D2-branes in the $r_{\rm M2} \rightarrow \infty $ limit,
we would like to fix the coupling constant (\ref{effective coupling on D2}).
Then, the decoupling limit of the D2-branes is given by 
\begin{align}
p^+ \rightarrow \infty,  \;\;\;\;\; \frac{R^2}{p^+}: {\rm fixed}.
\label{decoupling of D2}
\end{align}
The fixed quantity in $(\ref{decoupling of D2})$ measures 
the size of the M-circle for each fixed $p^+$, 
so that the M2-brane in 11-dimension 
is realized in the limit where $\frac{R^2}{p^+}$ becomes large.

The limit (\ref{decoupling of D2}) can be written in terms of the 
parameters of the matrix model. 
The D0-brane charge (the matrix size) 
$N$ is related to the light cone momentum by 
$p^+ = N/R$ and the gauge coupling of D0-branes is given as
$g^2 \sim R^3 l_p^{-6}$. 
Thus, the limit (\ref{decoupling of D2}) is translated to
\begin{align}
N \rightarrow \infty , \;\;\;\;\; \frac{g^2}{N}: \;{\rm fixed}. 
\end{align}
Then, the decoupling limit of M2-brane is given by sending 
$\frac{g^2}{N}$ to infinity.

Next, we consider the decoupling limit of the spherical M5-brane. 
The theory on NS5-branes is known as the little string theory. 
This theory is characterized by the string tension proportional to 
$1/l_{s}^2$. We can apply the above argument for D2-branes
to the little string theory. Here, the fixed quantity is replaced 
by the tension of the little string which is made 
dimensionless by using the radius of the M5-brane (\ref{radius of M5}):
\begin{align}
\frac{r_{\rm M5}^2}{l_s^2} 
\sim \frac{\tilde{R} r_{\rm M5}^2}{l_p^3} 
\sim \frac{R r_{\rm M5}}{\mu l_p^3}.
\end{align}
Thus, the decoupling limit of NS5-brane is given by
\begin{align}
p^+ \rightarrow \infty , \;\;\;\;\; R^4 p^+: {\rm fixed}.
\label{decoupling of NS5}
\end{align}
The M5-branes in 11-dimension are realized by further taking 
$R^4 p^+$ to be large.

In terms of the parameters of the matrix model,
the decoupling limit of NS5-brane (\ref{decoupling of NS5}) 
is translated into 
\begin{align}
N \rightarrow \infty , \;\;\;\;\; g^2N: \;{\rm fixed}. 
\label{M5 limit in PWMM}
\end{align}
This is just the 't Hooft limit of the matrix model\footnote{
In \cite{Ling:2006up}, a possible logarithmic correction to this limit 
was found.}.
The M5-brane limit corresponds to the strong coupling limit 
with respect to the 't Hooft coupling $g^2N$.

For multiple M5-branes, the radius of each M5-brane should become large to decouple from the gravity. 
Furthermore, if there are some stacks of M5-branes with different radii as shown in Fig.~\ref{young diagram}, the distances between the nearest stacks should 
also become large. The limit realizing this situation is such that the 
all radii become large with the same order. Since the radius of each M5-brane is proportional to a positive power of the light cone momentum, 
the decoupling limit for the multiple fivebranes
 should be given by (\ref{M5 limit in PWMM}) with $p_s^+/p_t^+ $ fixed for any $s,t=1,\cdots,\Lambda$. Thus, we find that the large-$N$ limit in (\ref{M5 limit in PWMM}) should be taken as in (\ref{limit M5}) in the case of the multiple 
M5-branes.

%%%%%%%%%%%%%%%%%%%%%%%%%%%%%%%%%%%%%%%%%%%%%%%%%%%%%%%%%%%%%%%%%%%%%%%%%%%%%%%
\section{The plane wave matrix model}
%%%%%%%%%%%%%%%%%%%%%%%%%%%%%%%%%%%%%%%%%%%%%%%%%%%%%%%%%%%%%%%%%%%%%%%%%%%%%%%
\label{The plane wave matrix model}
In this section, we review the plane wave matrix model (PWMM) 
\cite{Berenstein:2002jq}.

The Hamiltonian of PWMM is obtained by the matrix regularization of 
the Hamiltonian (\ref{HamM2}) of a single M2-brane
\cite{deWit:1988ig}.
In the matrix regularization, real functions on the world volume $f(\sigma^a)$ 
are linearly mapped to $N\times N$ Hermitian matrices, in such a way that
integrals and the Poisson algebra of functions 
are consistently mapped to traces and the commutator 
algebra of the corresponding matrices, respectively. 
Namely, under this mapping, we have 
\begin{align}
\frac{1}{4\pi }\int d^2 \sigma \rightarrow \frac{1}{N}{\rm Tr}, 
\;\;\;\;\;
\{ \;\; , \;\; \} \rightarrow \frac{-iN}{2}[\;\;, \;\; ].
\label{mapping integral}
\end{align}

For example, let us consider the case where
the spatial world volume is a unit sphere embedded in $R^3$.
The image of the embedding function $x^i$ which satisfies 
(\ref{embedding of s2}) is given by the $N$-dimensional irreducible 
representation of the $SU(2)$ generators,
\begin{align}
x_i \rightarrow \hat{x}_i= \frac{2}{N}L_i.
\label{hat x}
\end{align}
The normalization is chosen so that $\sum_i \hat{x}_i^2 =1_N$ holds 
in the large-$N$ limit. 
%Note that the algebra of functions on $S^2$ are generated by $\hat{x}_i$ 
%(i.e. the spherical harmonics can be represented as polynomials of 
%$\hat{x}_i$.).
One can check that (\ref{mapping integral}) 
is satisfied by (\ref{hat x}) for sufficiently large $N$.

By applying the matrix regularization to the Hamiltonian 
(\ref{HamM2}), we obtain the bosonic part of the 
Hamiltonian of PWMM as
\begin{align}
H=\frac{4\pi}{N}
{\rm Tr}
&\left[
\frac{4\pi }{2p^+}
\left( \left(\frac{N}{4\pi} \right)^2
P_A^2 -\frac{N^2T_{\rm M2}^2}{8}[X_A,X_B]^2
\right)
\right.
\nonumber\\
&\left.
+
\frac{p^+}{8\pi}
\left(
\frac{\mu^2}{9}X_i^2
+\frac{\mu^2}{36}X_a^2
\right)
+
\frac{i \mu N T_{\rm M2}}{12}
\epsilon^{ijk}
X_i [X_j,X_k]
\right].
\label{Ham of PWMM}
\end{align}
$P_A$ and $X_A\; (A=1,2,\cdots,9)$
are now $N \times N$ matrices, which correspond to the 
images of $P_A(\sigma^a)$ and $X_i(\sigma^a)$
in (\ref{HamM2}).
The constraint (\ref{constraint after gauge fixing}) is replaced by 
\begin{align}
[P_A, X^A] =0.
\label{constraint in Hamilton formalism}
\end{align}
In obtaining (\ref{Ham of PWMM}), we have also rescaled the momenta as
$P_A \rightarrow \left(\frac{N}{4\pi} \right) P_A$. 

The rescaled momenta correspond to the canonical momenta of $X^A$ in PWMM.
When one quantizes the theory of M2-brane (\ref{HamM2}), one has the canonical 
commutation relation\footnote{Here, the commutator represents the 
commutator of operators acting on the Fock space and this should 
not be confused with the commutator of $N\times N$ matrices.}, 
\begin{align}
[\hat{X}^A(\sigma ), \hat{ P}_B(\sigma')] = i \delta^A_B \delta^{(2)}
(\sigma - \sigma' ).
\end{align} 
Without the rescaling, according to (\ref{mapping integral}), this would be 
mapped to
\begin{align}
[\hat{X}^A_{ij}, \hat{ P}_{Bkl}] = i\frac{N}{4\pi} 
\delta^A_B \delta_{il}\delta_{jk}.
\end{align}
The rescaling just removes the factor $\frac{N}{4\pi}$ on the right-hand side
and makes $\hat{P}_{Aij}$ the canonically normalized momenta of 
$\hat{X}^A_{ij}$.
%\begin{align}
%[\hat{X}^A_{ij}, \hat{ P}_{Bkl}] = i \delta^A_B \delta_{il}\delta_{jk}.
%\end{align}
%The necessity of the rescaling can also be understood as follows.
%By applying the matrix regularization to the Lagrangian of 
%the M2-brane, one can obtain the Lagrangian of PWMM. 
%Then, one can switch to the Hamilton formalism through the 
%standard Legendre transformation of the matrix variables.
%Then, the canonical momenta defined by 
%$P_A=\frac{\partial L}{\partial \partial_t X^A} $ corresponds to the 
%rescaled momenta.

%In general, for any Hamiltonian $H(p,q)$ which is quadratic in the momentum 
%$p$, the classical equation of motion for $q$ is 
%invariant under the transformation
%$H(p,q) \rightarrow a H(p/a, q)$, where $a$ is a positive constant.
%In the matrix regularization, the hamiltonian is multiplied by a 
%constant $4\pi /N$ coming from the mapping of integrals
%(\ref{mapping integral}). 
%In order to preserve the form of the equation of motion, 
%we need the rescaling of $P_A$. 
%The rescaling of $P_A$ can also be understood 
%from  the basic requirement that 
%performing the matrix regularization should commute with
%switching from the Lagrangian formalism to the Hamiltonian formalism.

We consider vacua of PWMM. 
Noticing that the potential for $X_i$ forms a perfect square,
%as we saw in (\ref{perfect square M2}) before the matrix regularization, 
we find that the Hamiltonian is minimized when
\begin{align}
X^i = \frac{\mu p^+}{6\pi N T_{\rm M2}} L^i
\label{fuzzy sphere vacuum}
\end{align}
and the other fields are equal to zero.
Here, $L_i$ are $N$-dimensional representation matrices of the 
$SU(2)$ generators.
For any $N$-dimensional representation, 
(\ref{fuzzy sphere vacuum}) gives a vacuum of PWMM.
In particular, the representation is reducible in general 
and we can make an irreducible decomposition to express $L_i$ as in 
(\ref{irrdec}).
%where $L_i^{[N_5^{(s)}]}$ stand for the 
%$N_5^{(s)}$-dimensional irreducible representation matrices of 
%the $SU(2)$ generators and $N_2^{(s)}$ represents the multiplicity of 
%that representation.
%$\Lambda$ is the number of different irreducible representations
%which appear in the decomposition. 
With this decomposition,
the total matrix size can be written as $N = \sum_{s=1}^\Lambda N_2^{(s)}n_s$.
Thus, the vacua of PWMM are labeled by the discrete moduli parameters,
$\Lambda, N_2^{(s)}$ and $n_s$, which satisfy 
$N = \sum_{s=1}^\Lambda N_2^{(s)}n_s$. 

For later convenience, we introduce the action of PWMM.
We first rescale the matrices as
\begin{align}
Y^A = \frac{12 \pi N T_{\rm M2}}{\mu p^+} X^A.
\label{rescaling}
\end{align}
Then, the bosonic action of PWMM can be written in a simple form 
as\footnote{We have also rescaled 
the time coordinate appropriately.}
\begin{align}
S=\frac{1}{g^2}
\int dt 
{\rm Tr}
\left[
\frac{1}{2}(DY^A)^2 
-2 Y_i^2 -\frac{1}{2}Y_a^2 
+\frac{1}{4}[Y^A,Y^B]^2
-i\epsilon_{ijk} Y^i[Y^j,Y^k]
\right].
\label{action of PWMM}
\end{align}
Here, the coupling constant is related to the original parameters by
\begin{align}
g^2 = \frac{T_{\rm M2}^2}{2\pi } \left( \frac{12 \pi N}{\mu p^+} \right)^3
\label{redefining coupling}
\end{align}
and the covariant derivative is defined by
\begin{align}
DY^A = \frac{\partial}{\partial t} Y^A -i[A, Y^A].
\end{align}
The gauge field $A$ is introduced to take 
the constraint (\ref{constraint in Hamilton formalism}) into account.
In the $A=0$ gauge, the Gauss law constraint reproduces 
(\ref{constraint in Hamilton formalism}).

%%%%%%%%%%%%%%%%%%%%%%%%%%%%%%%%%%%%%%%%%%%%%%%%%%%%%%%%%%%%%%%%%%%%%%%%%%%%
\section{Spherical M5-branes from PWMM}
\label{Spherical M5-branes from PWMM}
%%%%%%%%%%%%%%%%%%%%%%%%%%%%%%%%%%%%%%%%%%%%%%%%%%%%%%%%%%%%%%%%%%%%%%%%%%%%
%For the vacuum given by the 
%$N$-dimensional single irreducible representation, 
%the radius of the fuzzy sphere is given by
%\begin{align}
%\frac{1}{N}{\rm Tr}X_i^2 = \frac{\mu^2(p^+)^2}
%{144\pi^2 T_{\rm M2}^2}\frac{N^2-1}{N^2}.
%\end{align}
%This approaches to $r_{\rm M2}^2$ in the large-$N$ limit.
%So this vacuum is expected to correspond to 
%the single spherical M2-brane in M-theory.

\subsection{Localization in PWMM}
We consider a complex scalar field in PWMM defined by
\begin{align}
\phi (t) = Y_3(t) + i(Y_8(t) \sin ( t) + Y_9(t) \cos (t)).
\end{align}
The real and imaginary parts of $\phi$ are given by 
an $SO(3)$ scalar and an $SO(6)$ scalar, respectively, 
up to the time dependent rotation. 
When one makes a double Wick-rotation for the time and $Y_9$ directions, 
one can construct (four) supercharges which leave $\phi$ invariant.
This allows us to exactly compute the expectation values of operators 
made of only $\phi$ by using the localization method
\cite{Pestun:2007rz}.
%\cite{Asano:2012zt,Asano:2014vba,Asano:2014eca}.

In order to perform the localization, one first needs to define the 
boundary conditions in the Euclidean time direction. 
Since we are interested in PWMM expanded around a fixed vacuum, 
the appropriate boundary condition is such that all the fields approach 
to the vacuum configuration as the Euclidean time goes to $\pm \infty$.
With this boundary condition, the path integral of PWMM defines the theory 
around the fixed background.

For the theory around the generic vacuum (\ref{irrdec}), 
the result of the localization obtained in 
\cite{Asano:2012zt,Asano:2014vba,Asano:2014eca}
is summarized below. See appendix \ref{Localization in PWMM} for the detail
of localization. 
We have the following equality:
\begin{align}
\langle \prod_I {\rm Tr}f_I(\phi (t_I)) \rangle
= 
\langle \prod_I {\rm Tr}f_I(2L_3+iM) \rangle_{MM},
\label{result of localization}
\end{align}
where $f_I(x)$ are arbitrary smooth functions, 
$2L_3$ is the vacuum configuration for $Y_3$. The matrix $M$ 
in (\ref{result of localization}) is an $N\times N$
constant Hermitian matrix which commutes with all of $L_a (a=1,2,3)$. 
For the representation given by (\ref{irrdec}), 
$M$ takes the form, 
\begin{align}
M= \bigoplus_{s=1}^{\Lambda}({\bf 1}_{n_s} \otimes M_s ),
\label{form of M}
\end{align}
where $M_s$ is an $N_2^{(s)}\times N_2^{(s)} $ Hermitian matrix.
The expectation value $\langle \cdots \rangle $ on the left-hand side of 
(\ref{result of localization}) is taken 
with respect to the original action 
%(\ref{action of PWMM}) 
of PWMM expanded around the background (\ref{irrdec}). 
On the other hand, the expectation value $\langle \cdots \rangle_{MM}$ on the 
right-hand side of (\ref{result of localization}) 
is taken with respect to the following matrix integral:
%%%%%%%%%%%%%%%%%%%%%%%%%%%%%%%%%%%%%
% caution introducing Z is a bit confusing.
%%%%%%%%%%%%%%%%%%%%%%%%%%%%%%%%%%%%%%
\begin{align}
Z= \int \prod_{s=1}^{\Lambda}
\prod_{i=1}^{N_2^{(s)}}
dq_{si} Z_{1-loop} e^{-\frac{2}{g^2}\sum_{s,i}n_s q_{si}^2},
\label{Z general}
\end{align}
where $q_{si} (i=1,2,\cdots, N_2^{(s)})$ are eigenvalues of ${M_s}$ 
and $Z_{1-loop}$ is the one-loop determinant, which arises in 
the 1-loop calculation of the localization. $Z_{1-loop}$ is given by 
\begin{align}
Z_{1-loop} = \prod_{s,t=1 }^{\Lambda} 
\prod_{J=|n_s-n_t|/2}^{(n_s+n_t)/2-1}
\prod_{i=1}^{N_2^{(s)}}
\prod_{j=1}^{N_2^{(t)}}{'}
\left[
\frac{\{(2J+2)^2+(q_{si}-q_{tj})^2\} \{(2J)^2 +(q_{si}-q_{tj})^2 \} }
{\{(2J+1)^2 +(q_{si}-q_{tj})^2\}^2}
\right]^{\frac{1}{2}}.
\label{z1-loop general}
\end{align}
The prime on the last product means that the second factor in the 
numerator with $s=t, J=0$ and $i=j$ is not included in the product.

Note that the right-hand side of (\ref{result of localization}) 
does not depend on the time coordinates $t_a$. 
So this relation implies that the correlator on the left-hand side
does not depend on time. 
This property can be understood from the SUSY Ward identity, as 
shown in \cite{Asano:2014vba}. 

We remark that, in the calculation of the localization,
some possible instanton corrections are neglected
\cite{Asano:2012zt,Asano:2014vba,Asano:2014eca}.
This corresponds to kink-like configurations in 
PWMM which connect two distinct vacua 
\cite{Yee:2003ge, Lin:2006tr, Bachas:2000dx,Kovacs:2015xha}.
However, the instanton amplitudes are bounded from below by 
$N_2/\lambda$ times
the difference of the quadratic Casimirs of the two vacua. 
Thus, in the decoupling limit of 
the M5-brane, this effect is suppressed.

\subsection{Coincident M5-branes from the simplest partition}
\label{Coincident M5-branes from the simplest partition}
To illustrate our computation, let us first consider the simplest 
partition with $\Lambda=1$, namely the vacuum with
\begin{align}
L_i = L_i^{[N_5]}\otimes 1_{N_2}.
\label{specific vacuum}
\end{align}
According to the proposal in \cite{Maldacena:2002rb}, 
this corresponds to $N_5$ coincident M5-branes.
In this case, the eigenvalue integral (\ref{Z general}) reduces to
a one matrix model:
\begin{align}
Z= \int \prod_i dq_i \prod_{J=0}^{N_5-1} \prod_{i>j}^{N_2}
\frac{\{(2J+2)^2+(q_{i}-q_{j})^2\} \{(2J)^2 +(q_{i}-q_{j})^2 \} }
{\{(2J+1)^2 +(q_{i}-q_{j})^2\}^2}
e^{-\frac{2N_5}{g^2}\sum_{i}q_i^2}.
\label{one matrix}
\end{align}

In the decoupling limit of the M5-brane, $N_2$ becomes infinity, 
so that the saddle point approximation is valid
in evaluating the eigenvalue integral (\ref{one matrix}). 
As usual, we introduce the eigenvalue distribution 
\begin{align}
\rho(q)= \frac{1}{N_2}\sum_{i=1}^{N_2}\delta (q-q_i),
\label{eigenvalue distribution def}
\end{align}
which is normalized as 
\begin{align}
\int_{-q_m}^{q_m}dq \rho(q) =1.
\label{normalization of rho}
\end{align}
Here, $q_m$ represents the range of the support of $\rho(x)$\footnote{
Note that $\rho$ has a single support, because 
the potential in (\ref{one matrix}) has a single well. }. 
Note that, we are interested in the decoupling limit of M5-brane where 
the 't Hooft coupling $\lambda:= g^2 N_2$ goes to infinity.
In this regime, the Gaussian attractive force 
of the eigenvalue integral (\ref{one matrix}) becomes weaker,
so that $q_m$ is expected to go to infinity.
If one considers the region where $q_m$ is very large compared to $N_5$,
one can reduce the saddle point equation of $\rho (x)$ to 
\begin{align}
\beta = \pi \rho(q) + \frac{2N_5}{\lambda }q^2 
- \int dq' \frac{2N_5}{(2N_5)^2+(q-q')^2}
\rho (q'),
\label{spe}
\end{align}
where 
%$\lambda=g^2 N_2$ is the 't Hooft coupling and 
$\beta$ is the Lagrange multiplier, which imposes the normalization 
(\ref{normalization of rho}).
See appendix \ref{The saddle point equation}
for the derivation of (\ref{spe})

In the M5-brane limit, the solution to the saddle point equation is given by
\begin{align}
\rho (q) = \frac{8^{3/4}}{3\pi \lambda^{1/4}}
\left[1-
\frac{q^2}{q_m^2}
\right]^{3/2}, \;\;\;  q_m=(8\lambda )^{1/4}, 
\;\;\; \beta = \frac{8^{1/2}N_5}{\lambda^{1/2}}.
\label{density solution}
\end{align}
See appendix 
\ref{Solving the saddle point equation}
for the derivation of this solution\footnote{
See also \cite{Asano:2014vba} for another derivation using the 
Fermi gas method.}.
Note that indeed $q_m$ becomes infinity as the 't Hooft coupling 
goes to infinity. 
%Note also that, in general,
%the saddle point approximation can be applied to the 't Hoof limit, where
%all terms of action become ${\cal O}(N^2)$. 
%Since the limit we consider is a further strong coupling limit,
%one may suspect that the solution may not be relevant to this coupling regime.
%However, one can also derive the solution (\ref{density solution}) by 
%using the Fermi gas method, which is valid in stronger coupling regime
%than the 't Hooft limit \cite{Asano:2014vba}. 
%Hence, the solution (\ref{density solution}) covers not only the 
%'t Hooft limit but also the strong coupling regime where M-theory is 
%considered to be realized.

By using this solution, 
we can compute correlation functions of $\phi$.
For example, 
\begin{align}
\frac{1}{N}\langle {\rm Tr}\phi^2 (0) \rangle 
&= 
\frac{1}{N}\langle {\rm Tr}Y_3^2 (0) \rangle - 
\frac{1}{N}\langle {\rm Tr}Y_9^2 (0) \rangle
= 
\frac{1}{N}\langle {\rm Tr}(2L_3+ iM)^2 \rangle_{MM}
\nonumber\\
&= 
\frac{1}{N}4{\rm Tr}(L_3^2) -
\frac{1}{N}\langle {\rm Tr}M^2 \rangle_{MM}
\nonumber\\
&= 
\frac{N_5^2-1}{12} - \int_{-q_m}^{q_m} dq q^2 \rho(q)
\nonumber\\
&=
\frac{N_5^2-1}{12} - \frac{\sqrt{8\lambda}}{6}.
\label{two point}
\end{align}
Note that the second term is much larger than the first term 
in the strong coupling regime with $N_5$ fixed. 
This originally comes from the fact that the eigenvalue 
distribution of $M$ spreads over the much wider region than 
the distribution of $L_3$ in the M5-brane limit.
This property is common for any correlation function of $\phi$, 
including the resolvent. In this regime, therefore, 
the imaginary part of $\phi$ is dominant and the real part is negligible. 
In other words, the spectrum of $\phi$ lies along the imaginary axis 
in this limit.

Assuming that the matrices $Y^A$ become mutually commuting in the 
decoupling limit, one may expect that this spectrum on the 
imaginary axis given by $\rho$ in (\ref{density solution}) 
could be identified with the eigenvalue distribution 
of one of the $SO(6)$ scalars. However, such identification 
would contradict with the discussion 
in \cite{Polchinski:1999br} by Polchinski. 
In \cite{Polchinski:1999br}, the BFSS matrix model
is considered and the trace of the square of the scalar fields $Y^A$ 
is shown to be bounded from below by $\lambda^{2/3}$ 
(in the notation used in this paper). 
And this conclusion is considered to hold also for PWMM 
if we assume the gauge/gravity correspondence: 
The dual geometry of PWMM reduces to the dual geometry 
of BFSS matrix model at a sufficiently large radius 
$r\gg {\cal O}(\lambda^{1/4})$ in the decoupling limit of M5-brane \cite{LM}.
On the other hand, if one assumes that $\rho$ in (\ref{density solution})
gives the eigenvalue distribution of one of the $SO(6)$ scalars in PWMM, 
this would give $\frac{1}{N}{\rm Tr }(Y^A)^2 ={\cal O}(\lambda^{1/2})$, which 
is smaller than the bound in the Polchinski's argument. 
Thus, this leads to a contradiction and 
the first assumption that $Y_A$ become commuting in the decoupling limit 
seems to be wrong\footnote{We thank J. Maldacena for suggesting this problem 
and the resolution using the time average which we will discuss below.}.

Apart from Polchinski's argument, we can find another reasoning for the 
above statement, based on the gauge/gravity correspondence.
The gravity dual \cite{LM} of PWMM has a typical scale $\lambda^{1/3}$, which 
is the string scale beyond which the supergravity approximation is not valid. 
It is natural to expect that the matrix elements of $Y^A$ contain information
of such typical scale on the gravity side, so that 
the scalar fields in PWMM have the typical value 
$\frac{1}{N}{\rm Tr }(Y^A)^2 ={\cal O}(\lambda^{2/3})$. 
Then, it is again suggested that the matrices are noncommuting 
even in the strongly coupled region.

The classical geometry of the supergravity and the M2/M5-branes are 
considered to be realized as the low energy moduli of these matrices. 
Roughly speaking, they will correspond to the low energy modes of the 
matrices and one needs to consider the low energy theory of the matrix model to find the classical geometric objects in M-theory.
The noncommuting modes, which produce the large value for 
$\frac{1}{N}{\rm Tr }(Y^A)^2 ={\cal O}(\lambda^{2/3})$, have a large 
excitation energy, so that these modes should be frozen and 
irrelevant in studying the low energy theory. 

Note that the complex field $\phi$ has the eigenvalue distribution of 
order of $\lambda^{1/4}$. This is much smaller than the typical value of
the noncommuting modes. From this fact, we find that 
$\phi$ is a good low energy field and the operators ${\rm Tr } \phi^n$ 
can be considered as operators in the low energy theory. 
This can also be understood from our formula (\ref{result of localization})
of the localization. The correlation functions of $\phi$ are 
independent of the time coordinates and hence are invariant under taking 
the time averages, which projects the operators to the low energy modes
(More specifically, one can eliminate the high energy modes by integrating 
over very short time intervals with length given by $1/C$, where $C$ 
is a constant much smaller than the typical energy scale 
for noncommuting modes but 
much larger than the energy scale for (\ref{density solution}).).
This means that the result of the localization (\ref{result of localization}) 
contains only the low energy modes.

As is discussed in \cite{Polchinski:1999br}, operators in 
the matrix model should be additively renormalized in the low energy theory, 
where the additive renormalization constants correspond 
to contributions from the high energy noncommuting modes.
However, such additive renormalization is not needed for ${\rm Tr } \phi^n$. 
In order for the eigenvalues of $\phi$ to be of ${\cal O}(\lambda^{1/4})$, 
the renormalization constants for $Y^A$ must cancel out in the correlators 
of $\phi$. 
For example, this can be seen in our computation in (\ref{two point}). 
Since the ${{\rm Tr} \phi^2}$ is given by the difference between 
${\rm Tr}(Y^3)^2$ and ${\rm Tr}(Y^9)^2$, 
the renormalization constants should cancel out. 

The statement that $\phi$ picks up the low energy moduli of the matrices 
is also supported by the earlier work on the gauge/gravity correspondence for 
PWMM. It was shown in \cite{Asano:2014vba,Asano:2014eca} that the field $\phi$ 
describes a system of moduli parameters on the gravity side, 
which is equivalent to a certain axially symmetric electrostatic system:
The charge densities of the electrostatic system, which determines the geometry
on the gravity side, were shown to be equivalent 
to the eigenvalue density of $\phi$.
%This fact suggests that $\phi$ should contain 
%the information of the moduli of PWMM.

From these observations, we claim that the spectrum of $\phi$ 
is identified with the low energy moduli of PWMM. 
Furthermore, we claim that the low energy moduli in PWMM are given by 
commuting matrices in the decoupling limit. 
This can be understood as follows.
Suppose that the moduli are given by noncommuting matrices and 
the theory on the M5-branes has some noncommutativity of the low energy 
moduli parameters as well as 
some length scale associated with the noncommutativity. 
The noncommutative length scale must be much smaller than 
the radius of the M5-brane, since otherwise the M5-brane
would not be localized along the radial direction 
due to the nonlocality caused by the noncommutativity
and hence would not be regarded as 1+5 dimensional object.
Then, let us consider the length scale $\lambda^{1/4}$
of the low energy moduli computed from the localization. 
This scale corresponds to the scale of the M5-brane radius 
if one takes the rescaling (\ref{rescaling}) into account.
Thus, the length scale of the low energy moduli must be 
much larger than the noncommutative scale. 
Therefore, even if the moduli have noncommutativity, 
this effect must be much smaller than the value of the moduli themselves in
the decoupling limit. 
Thus, we can ignore the noncommutativity 
and can regard the moduli as just commuting matrices.
Note that this conclusion is consistent with our result of the 
localization (\ref{result of localization}). Here, 
the moduli distribution is given by 
the distribution of $2L_3+iM$ in (\ref{one matrix}), and 
$L_3$ and $M$ are indeed mutually commuting variables.

The commutativity of the low energy moduli matrices might be general 
phenomena which occur in the strong coupling limit. 
As observed in \cite{Berenstein:2008eg}, in some matrix models 
with commutator interactions, 
commuting matrices indeed arise in the strong coupling limit.
A possible mechanism is as follows. For Yang-Mills type matrix models, 
one can rescale the matrices in such a way that
the coupling constant appears in front of each commutators.
In the strong coupling limit, in order to have a finite value of the action,
the values of commutators themselves 
must become small unless there is some cancellation with the kinetic terms. 
If this occurs, the matrices become commuting with each other. 
Though observing this phenomena directly 
in the current model is very difficult, this is very likely to occur 
in the low energy region, since in the low energy limit, 
the kinetic terms of the matrices are very small 
and there will be no chance to have a cancellation between 
the kinetic terms and the commutator terms.

Thus, we identify the real and imaginary parts of $\phi$ 
in the formula (\ref{result of localization})
with the low energy moduli for $Y^3$ and $Y^9$, respectively. 
In particular, $\rho$ in (\ref{density solution}) is identified with the 
moduli of $Y^9$. 
Recall that, in the decoupling limit of M5-brane, we have seen that 
the spectrum of $\phi$ becomes pure imaginary. 
Hence, with the suitable normalization of matrices 
(namely, going back to the original normalization in (\ref{rescaling})), 
one finds that the moduli of the $SO(6)$ scalar have 
a wide distribution while the moduli of the $SO(3)$ scalars 
collapse to the origin in the decoupling limit of the M5-brane. 

Now, let us consider the description of the spherical M5-brane.
%The shape of the spherical M5-brane in M-theory 
%can be characterized by the density function in $R^6$,
%\begin{align}
%\tilde{\rho} (r)= \frac{1}{V_5r_{\rm M5}^5}\delta(r-r_{\rm M5}),
%\label{spherical distribution}
%\end{align}
%where $r=\sqrt{x_a^2}$.
%The density function is normalized as 
%\begin{align}
%\int d^6x \tilde{\rho} (r)=1.
%\end{align}
We consider the $SO(6)$ symmetric uplift of the distribution 
\cite{Filev,Filev:2014jxa} of the moduli of a single $SO(6)$ scalar.
The uplifted distribution $\tilde{\rho}$ is defined as the solution of
\begin{align}
\int d^6 x \tilde{\rho}(r) x_9^{2n}
= \left(\frac{\mu p^+}{12 \pi N T_{\rm M2}}\right)^{2n}
\int^{q_m}_{-q_m}dq \rho(q) q^{2n},
\label{uplift}
\end{align}
for any $n$, where $r=\sqrt{x_a^2}$ is the distance from the origin. 
The normalization factor on the right-hand side is chosen so that 
$\tilde{\rho}$ represents a density function before the  
rescaling (\ref{rescaling}). 
For the density $\rho$ in (\ref{density solution}),
the unique solution to (\ref{uplift}) is
\begin{align}
\tilde{\rho} (r)= \frac{1}{V_5r_{0}^5}\delta(r-r_{0}).
\label{spherical distribution}
\end{align}
The radius $r_0$ is given by
\begin{align}
r_0= \left( \frac{\mu p^+}{6\pi^3 N_5 T_{\rm M5}} \right)^{1/4}.
\end{align}
For $N_5=1$, the shape of the density function of the $SO(6)$ moduli 
agrees with the shape of the spherical M5-brane. In particular, 
the radius shows a perfect agreement with the M5-brane: $r_0= r_{\rm M5}$.
Therefore, we conclude that the spherical M5-brane is indeed realized  
as the low energy moduli distribution of the $SO(6)$ scalar fields in PWMM.

For $N_5>1$, (\ref{spherical distribution}) should correspond to 
the radius of multiple coincident M5-branes.
The $N_5$-dependence of the radius agrees with 
the expected form in \cite{Maldacena:2002rb} based on the perturbative 
expansion in PWMM.

\subsection{Multiple M5-branes from generic partitions}
Let us generalize the above calculation to the case of 
the general partition (\ref{irrdec}). 
According to \cite{Maldacena:2002rb}, 
this corresponds to $\Lambda$ stacks of M5-branes with different radii
as shown in Fig.~{\ref{young diagram}}.
As we discussed in section~\ref{decoupling limits}, to make the M5-branes
decouple from the bulk gravity, we consider the limit
(\ref{M5 limit in PWMM}) such that 
the large-$N$ limit is taken as in (\ref{limit M5}). 

We introduce the eigenvalue distribution for $q_{si}$
in (\ref{Z general}) for each $s$ as
\begin{align}
\rho_s(q)= \sum_{i=1}^{N_2^{(s)}}\delta(q-q_{si})
\end{align}
and again assume that $\rho_s(q)$ has a single support $[-q_s, q_s]$. 
Note that, to simplify some expressions below, 
here we use the normalization
\begin{align}
\int_{-q_s}^{q_s} \rho_s(q) = N_2^{(s)},
\end{align}
which is different from the one we used in the previous subsection.
The saddle point equations for $\rho_s(q)$ 
can be derived in the same way as (\ref{spe}) and take the form,
\begin{align}
\rho_s(q)+\frac{1}{\pi} \sum_{t=1}^{\Lambda}
\int_{-q_t}^{q_t} du \left\{
\frac{|n_s-n_t|}{|n_s-n_t|^2+(u-q)^2}-
\frac{n_s+n_t}{(n_s+n_t)^2+(u-q)^2}
\right\} \rho_t (u)
= \frac{\mu_s}{\pi} -\frac{2n_s}{\pi g^2} q^2,
\label{saddle for general partition}
\end{align}
where $q\in [-q_s, q_s]$ and $s=1,2,\cdots,\Lambda$.
In appendix~\ref{Solving the saddle point equation 2}, 
we construct a solution to these equations in the 
decoupling limit.
The solution is given as 
\begin{align}
\hat{\rho}_s(q) = \frac{8^{3/4} \sum_{r=1}^s N_2^{(r)}}
{3\pi \lambda_s^{1/4} }\left[ 
1-\frac{q^2}{q^2_r} 
\right]^{\frac{3}{2}}, \;\;\;
q_s = \left(8\lambda_s \right)^{1/4}, \;\;\; 
\lambda_s := g^2\sum_{r=1}^s N_2^{(r)}, 
\label{solution for generic partition}
\end{align}
where $s=1,2,\cdots, \Lambda$ and $\hat{\rho}_s (q)$ are defined by
\begin{align}
\hat{\rho}_s (q) : = \sum_{r=1}^s \rho_r(q).
\end{align}

The variables $\hat{\rho}_s (q) (s=1,2, \cdots, \Lambda)$ 
have the following properties. 
First, $\hat{\rho}_s (q)$ is defined on the interval $[-q_s, q_s]$ and 
is normalized as 
\begin{align}
\int^{q_s}_{q_s}dq \hat{\rho}_s(q) = \sum_{r=1}^s N_2^{(s)}.
\end{align}
Note that $\sum_{r=1}^s N_2^{(s)}$ is proportional to the light cone momentum 
of the M5-brane in the $s$th stack (\ref{pplus for sth stack}).
Second,  $\hat{\rho}_s (q)$ naturally appear in evaluating the 
correlation functions of the complex field $\phi$. As we discussed in the 
previous subsection, in the decoupling limit of M5-branes, 
$L_i$ on the right-hand side in
(\ref{result of localization}) can be ignored. Then, we have for example,
\begin{align}
\langle {\rm Tr} \phi^{n} \rangle
&= i^n \langle {\rm Tr} M^n \rangle_{MM} = 
i^n   \sum_{s=1}^\Lambda \sum_{i=1}^{N_2^{(s)}}
n_s \langle q_{si}^n \rangle_{MM}
\nonumber\\
&= i^n \int dq q^n 
\left[
n_1 \rho_1(q)+n_2 \rho_2(q) + \cdots +
n_{\Lambda-1} \rho_{\Lambda-1}(q)+
n_\Lambda \rho_\Lambda(q)
\right]
\nonumber\\
&=i^n \int dq q^n 
\left[
(n_1-n_2) \hat{\rho}_1(q)+(n_2-n_3) \hat{\rho}_2(q) + \cdots +
(n_{\Lambda-1}-n_\Lambda) \hat{\rho}_{\Lambda-1}(q)
+ n_\Lambda \hat{\rho}_\Lambda(q)
\right].
\end{align}
Note that the coefficient $(n_s-n_{s+1})$ of $\hat{\rho}_s$ 
is just the number of M5-branes in the $s$th stack.
From these properties, $\hat{\rho}_s$ can be naturally 
identified with the density function for an M5-brane in 
the $s$th stack.

Obviously, the $SO(6)$ symmetric uplift of $\{ \hat{\rho}_s \}$ is 
given by $\Lambda$ stacks of the spherical shells. 
By taking the rescaling (\ref{rescaling}) into account, 
the $s$th stack has the radius 
\begin{align}
r_s = \frac{q_s \mu}{ 12 \pi R T_{\rm M2}} 
= \left(
\frac{\mu p^+_s}{6 \pi^3 T_{M_5}}
\right)^{1/4}, 
\label{radii of generic partition}
\end{align}
where $p^+_s$ is defined in (\ref{pplus for sth stack}). 
Thus, we have shown that, as shown in Fig.~\ref{young diagram},
the generic partition indeed describes 
concentric stacks of M5-branes with radii given by 
(\ref{radii of generic partition}).

\section{Spherical M2-branes from PWMM}
\label{Spherical M2-branes from PWMM}
So far, we considered the description of M5-branes in PWMM.
In this section, we apply the same analysis to the M2-brane limit.
Note that the emergence of the spherical D2-branes 
in the type IIA superstring theory can be understood even at the 
level of the classical action. However, it it still nontrivial whether we can 
observe the emergence in the strong coupling region of PWMM. 
Here, we study the emergence of M2-branes in 
the decoupling limit of the M2-branes.
In this section, we only consider the simplest partition 
(\ref{specific vacuum}) but the generalization is straightforward.

%We first take a limit where $N_5$ goes to infinity while 
%$\lambda/N_5$ is fixed. 
%This realizes D2-branes in type IIA superstring theory. 
%M2-branes are realized in the strong coupling region 
%where $\lambda/N_5$ is large.

In the M2-brane limit, where $N_5$ goes to infinity, 
the one-loop determinant of the 
eigenvalue integral (\ref{one matrix}) converges to the hyperbolic 
tangent function. Thus, we obtain
\begin{align}
Z= \int \prod_i dq_i \prod_{i>j}^{N_2}
\tanh^2 \left( \frac{\pi (q_i -q_j)}{2}
\right)
e^{-\frac{2N_5}{g^2}\sum_{i}q_i^2}.
\label{D2 partition function}
\end{align}
%
% caution order of q_m is manifest from this expression,
% when N_5/g^2 is fixed.
%
Note that the model depends only on $N_2$ and $g^2/N_5$.

The typical value of the eigenvalues of this model
should depend on $N_2$ and $g^2/N_5$.
Then, in the decoupling limit of M2-branes, the typical value 
is much smaller than $N_5$.
This implies that, in the result of the localization
(\ref{result of localization}), 
the eigenvalue distribution of $M$ is much narrower than 
that of $L_3$. 
Hence the spectrum of $\phi$ lies on the real axis in this limit.
This implies that the moduli of $Y^3$ are given by the classical 
vacuum configuration $2L_3$ while the moduli of $Y^9$ collapse to the origin. 
It is easy to see that the $SO(3)$ symmetric uplift of this configuration 
gives the two-sphere and the radius agrees with that of the spherical M2-brane
on the supergravity side for $N_2=1$.
Thus, we see that the spherical M2-brane is also realized as the moduli of 
$SO(3)$ scalars.

However, we should notice that, unlike the decoupling limit of M5-branes, 
the instanton corrections could contribute to the partition function 
in the M2-brane limit\footnote{This effect can naturally be 
understood as the instantons on the theory on D2-branes, 
which connect two vacua with different monopole charges
\cite{Maldacena:2002rb}.}.
If this is the case,
since the result of the localization does not include 
the instanton corrections \cite{Asano:2012zt,Asano:2014vba,Asano:2014eca}, 
our computation is not correct. 
Then, in order for our computation to make sense, we need to consider 
the limit where the number of M2-branes goes to infinity. 
In this limit, the instanton effects will be suppressed.
Thus, at least in the large-$N_2$ case, 
the result of the localization shows
the emergence of the spherical M2-branes in PWMM.

When $N_2$ is large, we can find an exact solution for
the eigenvalue distribution of (\ref{D2 partition function}) 
and can check that the 
typical value of the eigenvalues is 
proportional to $\left(\lambda/N_5\right)^{1/3}$ 
for large $\lambda/N_5$. See appendix
\ref{Eigenvalue distribution in the M2-brane limit}.

Of course, there is still a possibility that the instantons
do not affect our computation. For instance, 
this happens if there exists 
a fermionic zero mode at the saddle points of instantons 
in the localization computation.
This needs a further analysis of the localization saddles in PWMM.

\section{Summary and discussion}
\label{Summary}

In this paper, we tested a conjecture on the description of 
spherical M5-branes in the matrix model formulation of M-theory.
We considered the plane wave matrix model (PWMM), which is 
expected to describe the M-theory on the maximally supersymmetric 
11-dimensional plane wave geometry.

We first reviewed that, in the M-theory, 
there exist spherical M2- and M5- branes 
with zero light cone energy. 
These spherical branes are considered to be described as certain
vacuum states in PWMM.
This relation between the spherical branes and the vacua of PWMM 
is stated in \cite{Maldacena:2002rb}.
In particular, it is conjectured that a single spherical M5-brane 
corresponds to the trivial vacuum of PWMM.

Through a direct computation in PWMM using the localization, we 
showed that the spherical M2- and M5- branes are formed 
by the distribution of the moduli of $SO(3)$ and $SO(6)$ scalar fields, 
respectively. This result strongly supports the proposal 
in \cite{Maldacena:2002rb}. 

%In this argument, we assumed that the moduli fields in PWMM
%are given by commuting matrices in the relevant limits. 
%Though this is of course a reasonable assumption
%as we discussed in section 
%(\ref{Coincident M5-branes from the simplest partition}),
%the commutativity of moduli in PWMM is desirably 
%still a conjecture and should be proved rigorously. 
%
%
%, since the matrix model 
%is widely believed to describe a 
%classical theory of gravity in the low energy region\footnote{
%This is also suggested from general consideration on
%the strongly coupled region of matrix models. The matrices are likely to 
%become mutually commuting in the strong coupling region, 
%since one can always rescale the fields 
%in such a way that the coupling constant appears only in front of the 
%commutators in the action, so that only the commuting configurations 
%become domoinant \cite{Berenstein:2008eg}.}.
%
%However, the commutativity of moduli in PWMM is 
%still a conjecture and should be proved rigorously. 
%If this conjecture is false, the physical meaning of the uplifted 
%distribution derived in this paper might be explained by the methods in
%\cite{Berenstein:2012ts, Schneiderbauer:2016wub, Ishiki:2015saa}, which 
%can relate matrix configurations with smooth manifold. 

As we discussed in section 
\ref{Coincident M5-branes from the simplest partition},
we can assume that the moduli in PWMM are given by commuting matrices in 
the decoupling limit of the M5-branes.
Here, let us consider a possible effective theory of these commuting matrices 
in the decoupling limit.
We require the theory to have the $SO(6)$ symmetry 
and to be able to reproduce our result of the localization. 
For the case of coincident M5-branes, 
a possible solution to these requirements is given by 
a commuting matrix model with 6 matrices
defined by\footnote{
The same model was also considered in different contexts 
\cite{Filev:2014jxa,Berenstein:2005aa,Berenstein:2005jq}.} 
\begin{align}
 \hat S=N^2\left[
 \frac{m^2}{2}\int d^6\vec{y'}\,\hat\rho(\vec{y'})\, \vec{y}^2
 -\int d^6\vec{y}\, d^6\vec{y'}\, \hat\rho(\vec{y})\hat\rho(\vec{y'})\ln |\vec y-\vec{y'}|
 -\beta \left( \int d^6\vec{y'}\,\hat\rho(\vec{y'})-1 \right) 
 \right],
\label{low energy so6 matrix model}
\end{align}
where $\hat \rho$ is the distribution of moduli $y_i^a \; (a=4,5,\cdots,9, \; 
i=1,2, \cdots, N)$ for the $SO(6)$ scalars $Y^a$,
\begin{align}
\hat\rho(y)= \frac{1}{N}\sum_{i=1}^N \delta^{(6)}(\vec{y}_i -\vec{y}).
\end{align}
The Lagrange multiplier $\beta$ is introduced to impose the normalization 
condition on $\hat \rho$.
%\begin{align}
%\int d^6 y \hat\rho (y) =1.
%\label{normalization of hatrho}
%\end{align}
The second term in (\ref{low energy so6 matrix model})
is understood as the Vandermonde determinant 
$\prod_{i< j}|\vec{y}_i-\vec{y}_j|^2$ for the 
commuting matrices. 
We fix the parameter $m$ in (\ref{low energy so6 matrix model}) as
\begin{align}
m=(8\lambda)^{-\frac{1}{4}},
\label{eff mass}
\end{align}
so that the model reproduces the result of the localization below.
In the 't Hooft limit, the WKB approximation becomes exact. 
The saddle point equation is given by
\begin{align}
 \beta=\frac{m^2}{2}\vec{y}^2
 -\int d^6\vec{y'}\, \hat\rho(\vec{y'})\ln |\vec y-\vec{y'}|^2.
\end{align}
The solution to this equation is obtained in 
\cite{Berenstein:2005aa,Berenstein:2005jq,Filev:2014jxa} as
\begin{align}
 \hat\rho(\vec{y})
 % =\frac{2m^4}{\pi^3}\delta(1/m^2-\vec{y}^2)
 =\frac{1}{\pi^3|\vec{y}|^5}\delta(|\vec{y}|-\frac{1}{m}).
\end{align}
Note that, through the rescaling (\ref{rescaling}),
this is indeed  
equivalent to \eqref{spherical distribution} obtained from the localization.
Thus, the saddle point configuration of the commuting matrix model 
agrees with the configuration of the coincident spherical M5-branes. 
This agreement suggests that the commuting matrix model might be relevant to a
certain sector of the low energy theory of PWMM. 

It would be interesting to find more general commuting matrix model, which 
reproduces our result for the general partition.
In addition, we also need to investigate whether some 
low energy excitations can also be 
reproduced from the commuting matrix model or not.

Finding a good description of the low energy theory
should be one of the most important problem in understanding 
the description of the classical geometry in the matrix theory.
We hope that our result gives a clue to this problem.

\section*{Acknowledgments}
%%%%%%%%%%%%%%%%%%%%%%%%%%%%%%%%%%%%%%%%%%%%%%%%%%%%%%%%%%%%%
We thank D.~O'Connor, D.~Berenstein, J.~Maldacena, H.~Shimada and K.~Shimizu 
for valuable discussions.
Y.~A. was supported by a Dublin Institute for Advanced Studies scholarship.
The work of G. I. was supported, in part, 
by Program to Disseminate Tenure Tracking System, 
MEXT, Japan and by KAKENHI (16K17679). 
S. S. was supported by the MEXT-Supported Program for the
Strategic Research Foundation at Private Universities 
Topological Science (Grant No. S1511006). 
The work of S.~T. was supported by JSPS KAKENHI Grant Number 17K05414.

\appendix
\section{Localization in PWMM}
\label{Localization in PWMM}
In this appendix, we perform the localization and derive the 
formula (\ref{result of localization}). 
In this appendix, following the method in \cite{Pestun:2007rz}, 
we use a Lorentzian signature obtained by 
a double Wick rotation for the time-direction and the direction of 
one of the $SO(6)$ scalar fields.
To use some 10 dimensional notation, we relabel 
the $SO(3)$ scalar fields in PWMM as 
$(Y_1,Y_2,Y_3) \rightarrow (Y_2,Y_3,Y_4)$, the $SO(6)$ scalar fields as
$(Y_4,\cdots,Y_9) \rightarrow (Y_5,\cdots ,Y_{10})$
and the gauge field as $A\rightarrow Y_1$. The double Wick rotation 
is performed for the $Y_1$ and $Y_{10}$ directions and hence, the $Y_1$'s
direction is Euclidean and $Y_{10}$'s direction is Lorentzian. 
We also use $Y_{0}$ to express the scalar field in the Lorentzian signature, 
which is related to $Y_{10}$ by $Y_{0}=iY_{10}$.

\subsection{Off-shell supersymmetry of PWMM}
In the above notation, 
the full action of PWMM can be written in the 10-dimensional notation as
\begin{align}
S_{PW}&=\frac{1}{g^2}\int d\tau  \Tr\Bigl(
\frac{1}{4}\sum_{M,N=1}^{10}F_{MN}F^{MN}
+\frac{1}{2}\sum_{a=5}^{10}Y_aY^a
+\frac{i}{2}\sum_{M=1}^{10} \Psi \Gamma^M D_M \Psi
\Bigr),
\label{action of PWMM in 10D notation}
\end{align}
Here, $\Psi$ is the 10-dimensional Majorana Weyl spinor with 16 components
and we use the gamma matrices defined in \cite{Asano:2012zt}. 
We have also used the following notation:
\begin{align}
&F_{1M}=D_{1}Y_M=\partial_\tau Y_M-i[Y_1,Y_M] \quad (M\neq 1),\n
&F_{ij}=2\varepsilon_{ijk}Y_{k}-i[Y_{i},Y_{j}], \quad
F_{ia}=D_{i}Y_a=-i[Y_{i},Y_a], \quad
F_{ab}=-i[Y_a,Y_b], \n
&D_{1}\Psi=\partial_\tau \Psi-i[Y_1,\Psi], \quad
D_{i}\Psi=\frac{1}{4}\varepsilon_{ijk}\Gamma^{jk}\Psi-i[Y_{i},\Psi], \quad
D_a\Psi=-i[Y_a,\Psi],
\label{F in PWMM}
\end{align}
where $i,j,k=2,3,4$ and $a,b=5,6,\cdots,10$.
In order to realize the off-shell supersymmetries, we further add 
seven auxiliary fields 
\begin{align}
-\frac{1}{g^2} \int d\tau \frac{1}{2} \sum_{I=1}^7 {\rm Tr}K_IK_I
\label{KK action}
\end{align}
to the action (\ref{action of PWMM in 10D notation}).
Under the Wick rotation, $K_I$ shall become anti-Hermitian, so that 
(\ref{KK action}) becomes positive definite in the Euclidean signature.

The theory has the off-shell supersymmetry,
\begin{align}
&\delta _sY_M=-i\Psi \Gamma _M\epsilon ,\nonumber \\
&\delta _s\Psi =\frac{1}{2}F_{MN}\Gamma ^{MN}\epsilon 
-Y_a\tilde{\Gamma}^{a}\Gamma^{19}\epsilon 
+K^I\nu _I,\nonumber \\
&\delta _sK_I=i\nu_I\Gamma ^MD_M\Psi.
\label{offshellsusy}
\end{align}
See \cite{Asano:2012zt} for the definition of $\tilde{\Gamma}^a$.
The parameter $\epsilon$ has to satisfy the Killing spinor equation of PWMM
and the closure of the supersymmetry requires 
$\nu_I$ to satisfy 
\begin{align}
&\epsilon \Gamma ^M\nu _I=0,\nonumber \\
&\frac{1}{2}(\epsilon \Gamma _N\epsilon )\tilde \Gamma ^N_{\alpha \beta}=\nu ^I_\alpha \nu ^I_\beta +\epsilon _\alpha \epsilon _\beta ,\nonumber \\
&\nu _I\Gamma ^M\nu _J=\delta _{IJ}\epsilon \Gamma ^M\epsilon .
\label{closedness}
\end{align}
The following spinors give a solution to these conditions:
\begin{align}
&\epsilon=
e^{\frac{\tau}{2}\Gamma ^{09}}e^{-\frac{\pi}{4}\Gamma ^{49}}
\begin{pmatrix}
\eta _1\\
0\\
0\\
0
\end{pmatrix}, \;\;\;\; 
\nu _I=\sqrt{2}
e^{\frac{\tau}{2}\Gamma ^{09}}e^{-\frac{\pi}{4}\Gamma ^{49}}\Gamma ^{I8}
\begin{pmatrix}
\eta _1\\
0\\
0\\
0
\end{pmatrix},
\label{CKS in PWMM}
\end{align}
where $\eta_1$ is any 4-component constant vector.
We use $\eta_1 =(1,0,0,0)$ in the following computation.

\subsection{Saddle point of the localization}
To perform the localization, we add an exact term $t \delta_s V$ 
to the action, where 
\begin{align}
V = \int d\tau {\rm Tr} \Psi \overline{\delta _s\Psi }.
\end{align}
After some calculation, one can find that the bosonic part of 
$\delta _sV$ is calculated to be
\begin{align}
\delta _sV
%\nonumber \\
&\sim
-e^{\tau}(D_1Y_0+Y_0-e^{-\tau}K_5)^2
-e^{-\tau}(D_1Y_0-Y_0+e^{\tau}K_5)^2
-2c \sum_{i=2}^4(D_{i}Y_0)^2\nonumber \\ 
&\quad 
-2c \sum_{ I\neq 5}(K^{I})^2
+2c (D_4Y_9)^2
+2c [Y_0,Y_9]^2
+2c \sum_{a=5}^8 [Y_0,Y_{a}]^2
+\mathcal{S}
\nonumber \\
&\quad 
+4\sum _{a=1}^{3}\left[
e^{-\tau }
\left\{ F_{a4}^+-\frac{1}{2}D_a(e^{\tau }Y_9)+F_{a+4,8}^+\right\} ^2
+e^{\tau }
\left\{ F_{a4}^-+\frac{1}{2}D_a(e^{-\tau }Y_9)-F_{a+4,8}^-\right\} ^2
\right],
%+\sum_{\nu =1}^4\sum_{a,b=1}^{4}\Big[
%e^{-\tau }(D_a(J_\nu )_{5-a,5-b}(e^\tau X_{b+4})+(-)^{\nu +1}F_{9,9-\nu})^2\nonumber \\
%&\qquad \qquad \qquad \qquad 
%+e^{\tau }(D_{a}(\Sigma J_\nu )_{5-a,5-b}(e^{-\tau}X_{b+4})-(-)^{\nu +1}F_{9,9-\nu})^2
%\Big] , %$\Sigma =\mathrm{diag}(-1,-1,-1,+1)$,
\label{pos-def}
\end{align}
%\begin{align}
%&4e^{-\tau }
%\Bigg[ \left\{ F_{14}^+-\frac{e^{\tau }}{2}(D_1X_9+X_9)+\frac{F_{58}-F_{67}}{2}\right\} ^2\nonumber \\
%&\quad +\left\{ F_{24}^+-\frac{e^{\tau }}{2}(D_2X_9)+\frac{F_{57}+F_{68}}{2}\right\} ^2 
%+\left\{ F_{34}^+-\frac{e^{\tau }}{2}(D_3X_9)+\frac{F_{78}-F_{56}}{2}\right\} ^2 \Bigg]\nonumber \\
%+&4e^{\tau }
%\Bigg[ \left\{ F_{14}^-+\frac{e^{-\tau }}{2}(D_1X_9-X_9)-\frac{F_{58}-F_{67}}{2}\right\} ^2\nonumber \\
%&\quad +\left\{ F_{24}^-+\frac{e^{-\tau }}{2}(D_2X_9)-\frac{F_{57}+F_{68}}{2}\right\} ^2
%+\left\{ F_{34}^-+\frac{e^{-\tau }}{2}(D_3X_9)-\frac{F_{78}-F_{56}}{2}\right\} ^2 \Bigg] ,
%\label{pos-def 4}
%\end{align}
where $c:=\cosh \tau$ and ${\mathcal S}$ is defined by
\begin{align}
\mathcal{S}&=e^{\tau }(Y_5+D_1Y_5+D_2Y_6+D_3Y_7+D_4Y_8+e^{-\tau }F_{98})^2\nonumber \\
&\quad +e^{-\tau }(Y_5-D_1Y_5-D_2Y_6-D_3Y_7+D_4Y_8-e^{\tau }F_{98})^2\nonumber \\
&\quad +e^{\tau }(Y_6+D_1Y_6-D_2Y_5+D_3Y_8-D_4Y_7-e^{-\tau }F_{97})^2\nonumber \\
&\quad +e^{-\tau }(Y_6-D_1Y_6+D_2Y_5-D_3Y_8-D_4Y_7+e^{\tau }F_{97})^2\nonumber \\
&\quad +e^{\tau }(Y_7+D_1Y_7-D_2Y_8-D_3Y_5+D_4Y_6+e^{-\tau }F_{96})^2\nonumber \\
&\quad +e^{-\tau }(Y_7-D_1Y_7+D_2Y_8+D_3Y_5+D_4Y_6-e^{\tau }F_{96})^2\nonumber \\
&\quad +e^{\tau }(Y_8+D_1Y_8+D_2Y_7-D_3Y_6-D_4Y_5-e^{-\tau }F_{95})^2\nonumber \\
&\quad +e^{-\tau }(Y_8-D_1Y_8-D_2Y_7+D_3Y_6-D_4Y_5+e^{\tau }F_{95})^2.
\label{pos-def 5}
\end{align}
The derivatives $D_{M}$ are defined in (\ref{F in PWMM}).
$F_{ab}^{\pm}$ stands for the selfdual and anti-selfdual part of 
$F_{ab}^{\pm}$ in the subspace $a,b=1,2,3,4$ or $a,b=5,6,7,8$.
After the Wick rotation, $Y_0=iY_{10}$ and $K_i=iK_i^{(E)} (i=1,2,\cdots,7)$,
the bosonic part $\delta _sV|_{bos}$
becomes a sum of positive-definite terms.

We consider the theory around a fixed vacuum (\ref{irrdec}). 
Then, we impose the boundary condition such that all fields approach 
to the vacuum configuration at $\tau \rightarrow \pm \infty $.
Then, taking the temporal gauge $Y_1=0$, we find that
the saddle point configuration is given by
\begin{align}
&\hat Y_{10}=\frac{M}{c},\quad  
\hat K_5^{(E)}=\frac{M}{c^2}, \quad
\hat{Y}_{i}= -2L_{i-1} \; (i=2,3,4),
\label{QV=0}
\end{align}
where all the other fields are zero.
Here, $2L_{i}$ $ (i=1,2,3)$ are the vacuum configuration and $M$ 
is a constant Hermitian matrix, which commutes with all of $L_i$.
For the vacuum of the form (\ref{irrdec}), $M$ takes the form 
(\ref{form of M}).

It is easy to see that the gaussian part in (\ref{Z general}) is obtained by 
substituting the saddle point configuration to the classical action $S_{PW}$.
The remaining part $Z_{1-loop}$ in (\ref{Z general}) is obtained by the
1-loop calculation around the saddle point.

\subsection{Ghost fields}
We introduce the collective notation,
\begin{align}
X = \left(
\begin{array}{c}
Y_A \\
(\epsilon \epsilon)\Upsilon_I
\end{array}
\right), \;\;\; 
X' = \left(
\begin{array}{c}
-i(\epsilon \epsilon)\Psi_A \\
H_I
\end{array}
\right),
\end{align}
where $\Upsilon_I, H_I \;(I=1,2,\cdots,7)$ and
$\Psi_A\; (A=1,\cdots,9)$ are defined below.
Since $\{\Gamma^A\epsilon, \nu_I | A=1,\cdots,9, \; 
I=1,\cdots,7\}$ gives an orthogonal basis for 16 
component spinors, $\Psi$ can be expanded as 
\begin{align}
\Psi = \Psi_A \Gamma^A \epsilon + \Upsilon_I \nu^I.
\end{align}
$\Psi_A$ and $\Upsilon_I$ are introduced as the 
coefficients of this expansion. $H_I$ are defined as
\begin{align}
H_I = (\epsilon \epsilon)K_I + 2\nu_I \tilde{\epsilon}
Y_0 + \nu_I \left(
\frac{1}{2}\sum_{A,B=1}^9F_{AB}\Gamma^{AB}\epsilon
-2\sum_{a=5}^9 X_a \Gamma^a \tilde{\epsilon}
\right),
\end{align}
where $\tilde{\epsilon}=\frac{1}{2}\Gamma^{19}\epsilon$.
We also define 
\begin{align}
\phi = Y_0 \cosh \tau -Y_4+ Y_9 \sinh \tau .
\end{align}
Then, the supersymmetry can be written as
\begin{align} 
\delta_s X = X', \;\; 
\delta_s X' = -i(\delta_\phi +\delta_{U(1)} )X, \;\;
\delta_s \phi =0,
\end{align}
where $\delta_{\phi}$ is a gauge transformation with 
the parameter given by $\phi$ and $\delta_{U(1)}$ is 
a diagonal $U(1)$ transformation of the 
$SO(3)\times SO(6)$ symmetry.
This shows that $X$ and $X'$ forms a doublet while 
$\phi$ is a singlet under the supersymmetry.

We also introduce the ghost fields,
$(C,C_0,\tilde{C},\tilde{C}_0,b,b_0,a_0,\tilde{a}_0)$, 
where $(b,b_0,a_0,\tilde{a}_0)$ are bosonic and
$(C,\tilde{C},C_0,\tilde{C}_0)$ are fermionic fields. 
The fields with subscript $0$ shall 
contain only zero modes for both $\tau$ direction 
and the fuzzy sphere directions.
They are defined through the following 
BRS transformations,
\begin{align}
&\delta_{B}X=[X,C],\;\;\;\;
\delta_{B}X'=[X',C],
\nonumber\\
&\delta_{B}C=a_0-C^2,\;\;
\delta_{B}\phi=[\phi,C],
\nonumber\\
&\delta_{B}\tilde{C}=b,\;\;\;\;\;\;\;\;\;\;\;\;
\delta_{B}b=[\tilde{C},a_0],
\nonumber\\
&\delta_{B}\tilde{a}_0=i\tilde{C}_0,\;\;\;\;\;\;\;\;
\delta_{B}\tilde{C}_0=-i[\tilde{a}_0,a_0],
\nonumber\\
&\delta_{B}b_0=iC_0,\;\;\;\;\;\;\;\;
\delta_{B}C_0=-i[b_0,a_0], \;\;\;\; \delta_B a_0=0.
\label{BRS transformation}
\end{align}
The commutator in the above equation
shall express the anti-commutator for fermionic 
variables. The square of $\delta_B$ is a gauge 
transformation with parameter $a_0$,
\begin{align}
\delta_B^2=[\;\; , a_0].
\end{align}

We define the supersymmetry transformation of the ghost fields
as
\begin{align}
\delta_s C = \phi,  \;\;\; \delta_s({\rm the \; other \; ghosts})=0.
\end{align}
Then $Q=\delta_s+\delta_B$ has the following action:
\begin{align}
&QX=X'+[X,C],\;\;\;\;
QX'=-i(\delta_{\phi}+\delta_{U(1)})X+[X',C],
\nonumber\\
&QC=\phi+a_0-C^2,\;\;\;\;
Q\phi=[\phi,C],
\nonumber\\
&Q\tilde{C}=b,\;\;\;\;\;\;\;\;\;\;\;\;\;\;\;\;\;\;\;\;
Qb=[\tilde{C},a_0],
\nonumber\\
&Q\tilde{a}_0=i\tilde{C}_0,\;\;\;\;\;\;\;\;\;\;\;\;\;\;\;\;
Q\tilde{C}_0=-i[\tilde{a}_0,a_0],
\nonumber\\
&Qb_0=iC_0,\;\;\;\;\;\;\;\;\;\;\;\;\;\;\;\;
QC_0=-i[b_0,a_0], \;\;\;\; Q a_0=0.
\label{Q transformation}
\end{align}
One can easily show that $Q^2$ is given as
\begin{align}
Q^2 =-i \delta_{U(1)}+[\;\;, a_0].
\label{Q square}
\end{align}

The gauge-fixing and ghost actions are defined by
\begin{align}
S_{gh}&=\int d\tau \; Q
{\rm Tr} \left[ 
i \tilde{C}\left( F+b_0  \right)
+C \tilde{a}_0 
\right],
\label{ghost action}
\end{align}
where $F$ corresponds to the gauge fixing condition.
We use
\begin{align}
F=\sum_{a=1}^4\hat{D}_a\left(\frac{1}{\cosh \tau} Y_a\right)
\end{align}
for our computation, where
the background covariant derivative $\hat{D}_a$ is defined by
\begin{align}
\hat{D}_a X :=-i [\hat{Y}_a, X ] \;\; (a=1,2,3,4).
\end{align}
Here, $\hat{Y}_1 = i\frac{\partial}{\partial \tau}$ and $\hat{Y}_i (i=2,3,4)$
are the vacuum configuration of $Y_i$.

\subsection{1-loop determinants}
Let us perform the 1-loop calculation 
around the saddle point (\ref{QV=0}).
We first redefine the fields as
\begin{align}
\tilde{X}':=X'+[X,C], \;\;\;
\tilde{\phi}&:=2\phi +a_0-C^2,
\end{align}
and divide the fields to four groups as
\begin{align}
Z_0&=(Y_{A},\tilde{a}_0, b_0), \;\; 
Z_1=(\Upsilon_I,C,\tilde{C}),
\nonumber\\
Z'_0&=(\tilde{\Psi}_{A},\tilde{C}_0,C_0), \;\; 
Z'_1=(\tilde{H}_I,\tilde{\phi},b).
\end{align}
They form doublets under the action of $Q$ as
\begin{align}
&QZ_i =Z'_i,  \;\;\; QZ'_i = RZ_i, \;\;\; (i=0,1)
\end{align}
where $R:=Q^2$ is given by the sum of the $U(1)$ and gauge transformations 
as shown in (\ref{Q square}).

Then we expand the full action $S_{PW}+tQ(V+V_{gh})$ 
around the saddle point configuration (\ref{QV=0}) as
$Z_i \rightarrow \hat{Z}_i+Z_i $ and 
$Z'_i \rightarrow \hat{Z'}_i+Z'_i $. 
Then the quadratic part of the fluctuations in $V+V_{gh}$
is schematically written as 
\begin{align}
V^{(2)}=(Z'_0,Z_1)
\left(
\begin{array}{cc}
D_{00} & D_{01} \\
D_{10} & D_{11} \\
\end{array} 
\right)
\left(
\begin{array}{c}
Z_0 \\
Z_1' \\
\end{array} 
\right),
\end{align}
where $D_{ij} (i,j=0,1)$ are some linear differential operators.
Thus, the quadratic part of the action takes the form
\begin{align}
QV^{(2)}= (RZ_0,Z'_1)
\left(
\begin{array}{cc}
D_{00} & D_{01} \\
D_{10} & D_{11} \\
\end{array} 
\right)
\left(
\begin{array}{c}
Z_0 \\
Z_1' \\
\end{array} 
\right)+
(Z_0',Z_1)
\left(
\begin{array}{cc}
D_{00} & D_{01} \\
D_{10} & D_{11} \\
\end{array} 
\right)
\left(
\begin{array}{c}
Z'_0 \\
RZ_1 \\
\end{array} 
\right).
\label{quadratic action}
\end{align}
Hence, the one-loop integral produces the determinants,
\begin{align}
Z_{\rm 1-loop}=\left(\frac{{\rm det}_{V_{Z_1}}R}{{\rm det}_{V_{Z_0}}R} \right)
^{\frac{1}{2}}.
\label{detR over detR}
\end{align}
Here, the determinants should be taken in the appropriate 
functional spaces of the fluctuations.
Recall that we adopted the boundary condition such that all fields
go to the vacuum configuration as $\tau \rightarrow \infty$. 
This implies that the fluctuations should vanish at infinities.

Note that $D_{10}$ is a linear map from 
$V_{Z_0}$ to $V_{Z_1}$ and commutes with $R$.
Then the determinants in (\ref{detR over detR}) 
cancel between ${\rm Im}D_{10}\subset V_{Z_1}$ 
and ${\rm Im}D_{10}^* \subset V_{Z_0}$, where
$D_{10}^*$ is the adjoint of $D_{10}$.
Hence, the 1-loop determinant reduces to 
\begin{align}
Z_{\rm 1-loop}=\left(\frac{{\rm det}_{{\rm coker}D_{10}}R}
{{\rm det}_{{\rm ker}D_{10}}R} 
\right)
^{\frac{1}{2}}.
\end{align}
Furthermore, since $R$ and $D_{10}$ commute, 
the kernel and the cokernel are given by
direct sums of the eigenspaces of $R$. Thus, we can express
the 1-loop determinant as
\begin{align}
Z_{\rm 1-loop}= 
\prod_{i}r_i^{({\rm dim}V'_{r_i}-{\rm dim}V_{r_i})/2},
\label{z pert}
\end{align}
where $V_{r_i}$ and $V'_{r_i}$ are the restrictions of 
the kernel and the cokernel to the eigenspace of $R$ with eigenvalue $r_i$, 
respectively.
Therefore, the remaining task is to evaluate $r_i$ and the  
index ${\rm dim}V'_{r_i}-{\rm dim}V_{r_i}$ in each eigenspace. 

By integrating the ghost field $\tilde{a}_0$, we obtain the 
constraint $a_0=-2\phi$. At the saddle point, this is 
equal to $-2iM+4L_4$. Thus, $r_i$ is given by the sum of eigenvalues of
$[-2iM+4L_4, \;\;]$ and the diagonal $U(1)$ charge.

By studying the structure of $D_{10}$ for each 
supersymmetry multiplet, we can easily compute the index. 
The result is as follows \cite{Asano:2012zt}. 
The contribution from the hypermultiplet, 
which contains $Y_5,\cdots, Y_8$, is given by 
\begin{align}
\prod_{s,t=1}^{\Lambda}\prod_{J=|n_s-n_t|/2}^{(n_s+n_t)/2-1}
\prod_{i=1}^{N_2^{(s)}}\prod_{j=1}^{N_2^{(t)}}
\frac{1}{(2J+1)^2+(q_{si}-q_{tj})^2}.
\label{z hyper}
\end{align}
The contribution from the vector multiplet, which contains 
$Y_1,\cdots, Y_4, Y_9$, is given by
\begin{align}
&\prod_{s,t=1}^{\Lambda}
\prod_{\substack{J=|n_s-n_t|/2 \\J\neq 0}}^{(n_s+n_t)/2-1}
\prod_{i=1}^{N_2^{(s)}}\prod_{j=1}^{N_2^{(t)}}
\{(2J)^2+(q_{si}-q_{tj})^2 \}^{1/2}
\nonumber\\
\times & \prod_{s,t=1}^{\Lambda}\prod_{J=|n_s-n_t|/2}^{(n_s+n_t)/2-1}
\prod_{i=1}^{N_2^{(s)}}\prod_{j=1}^{N_2^{(t)}}
\{(2J+2)^2+(q_{si}-q_{tj})^2 \}^{1/2}.
\label{z vector}
\end{align}
Combining these contributions with the Vandermonde determinant 
for diagonalizing $M$, we obtain the 1-loop determinant 
(\ref{z1-loop general}). See below for the derivation of these 
1-loop determinants.

\subsection{Derivation of 1-loop determinants}

The relevant part of the action is given by $Z_1D_{10}Z_0$.
In terms of the component fields, this can be written explicitly as
\begin{align}
&2s_i \Upsilon_i+i\tilde{C}(F+b_0)+C\tilde{a}_0
\nonumber\\
&-\frac{i}{\epsilon \epsilon}
\left( 
\delta_{U(1)}Y_{A}
-2i[\hat{Y}_{A},v^4Y_4+v^9Y_9]
-i[Y_{A},-2iM+v^4\hat{Y}_4]
\right)
[\hat{Y}_{A},C],
\label{linearized action}
\end{align} 
where
\begin{align}
s_i :=\nu _i\left( \frac{1}{2}\sum_{A,B=1}^9F_{AB}\Gamma ^{AB}\epsilon 
-2\sum_{a=5}^9X_{a}\Gamma ^{a}\tilde \epsilon \right).
\end{align}

Note that the fields in the hypermultiplet, 
$\{ (Y_{m},\Upsilon_i) | m=5,8,7,8, i=1,2,3,4 \}$, 
decouple from the fields in the vector multiplet in 
(\ref{linearized action}).
Hence, the index has two independent contributions from these two sectors.

\subsubsection*{Index theorem in 1-dimension}
For the computation of the 1-loop determinant, 
the index theorem in 1-dimension is very useful, which we will describe below.

The setup is as follows. We consider the set of all $n$-dimensional 
vector valued smooth functions on $R$ vanishing at infinity, 
$S:=\{f: R \rightarrow  C^n | 
\lim_{\tau \rightarrow \pm \infty}f(\tau)= 0 \}$.
Let us introduce a linear differential operator $D$ on $S$ as
\begin{align}
Df(\tau):=\frac{\partial f}{\partial \tau}(\tau) +(A\cdot f)(\tau),
\label{df=0}
\end{align}
where $f \in S$ and $A: R \rightarrow M_n(C)$.
$A\cdot f$ is just the standard action of matrices,
$(A\cdot f)_i(\tau):=A_{ij}(\tau)f_j(\tau)$.
For the computation of the 1-loop determinant, 
we only consider the case where $A$ is bounded at both infinities as
$\lim_{\tau\rightarrow \pm \infty} A_{ij}(\tau)<\infty$ $(i,j=1,\cdots,n)$ 
and $A(\tau)$ is diagonalizable as 
\begin{align}
V^{-1}(\tau)A(\tau)V(\tau)=A_d(\tau)
:={\rm diag}(\lambda_1(\tau), \cdots, \lambda_n(\tau)).
\label{diagonalize}
\end{align}
%by a certain $V \in \Gamma(E)$, where 
%$E$ is $SL(n,C)$ bundle on $R$ and $\Gamma(E)$ is the set of 
%all smooth sections of $E$.
%$V$ is not uniquely determined by (\ref{diagonalize}). 
As $A$ is bounded, both of $\lim_{\tau\rightarrow \pm \infty}A_{ij}(\tau)$
and $\lim_{\tau\rightarrow \pm \infty}\lambda_i(\tau)$ are some constants. 
Then,
$\lim_{\tau\rightarrow \pm \infty}V(\tau)$ are also constant matrices.

The 1-dimensional index theorem follows from the fact that 
the number of positive and negative eigenvalues of $A$ 
at both infinities determines the index of $D$.
The essential statement of the index theorem is that 
if the $k$ $(1\leq k \leq n)$ eigenvalues in 
(\ref{diagonalize}) satisfy both 
\begin{align}
\lim_{\tau \rightarrow \infty} {\rm Re}\lambda_i(\tau)>0 \;\;\; {\rm and} 
\;\;\;
\lim_{\tau \rightarrow -\infty} {\rm Re}\lambda_i(\tau)<0 
\label{hantei}
\end{align}
and the remaining $n-k$ eigenvalues do not, 
then, we have
\begin{align}
{\rm dim}({\rm ker} D) = k.
\label{dimker}
\end{align}
This relation can be shown as follows. 
Note that $D$ is covariant under $A \rightarrow 
U^{-1}AU+U^{-1}\partial U$. 
Consider the gauge transformation such that
\begin{align}
U^{-1}AU+U^{-1}\partial U =A_d.
\end{align}
Such $U$ can be expressed as
$U(\tau)=[P\exp(-\int^\tau A)]\exp(\int^\tau A_d)$, 
where $P$ denotes the path ordering.
The general solution to the differential equation $Df=0$ is then given by
\begin{align}
f(\tau)=U(\tau)\exp\left(-\int^{\tau}_0 A_d(\tau')d\tau' \right)f_0,
\label{formal solution}
\end{align}
where $f_0$ is a constant vector.
In order to be a solution in the space of $S$, 
(\ref{formal solution}) has to vanish at both infinities. 
Here, let us consider the condition (\ref{hantei}).
When $k$ of $\lambda_i$'s satisfy (\ref{hantei}),
only $k$ components of $f_0$ can be nonzero to satisfy the 
boundary conditions. This implies (\ref{dimker}).

Of course, the similar equation to (\ref{dimker}) holds for 
the adjoint operator $D^\dagger$. By combining this with (\ref{dimker}),
we obtain the index theorem in 1-dimension, which states that 
the index of $D$ is completely determined by the behavior of $A$ 
at infinities. 

\subsubsection*{Hypermultiplet}
Let us consider the hypermultiplet. 
We use complex combinations,
\begin{align}
W_1=Y_5+iY_8, \;\;\; W_2=Y_6+iY_7.
\end{align}
We can read off the action of $D_{10}$ from (\ref{linearized action}).
If $(W_1, W_2)$ is an element of $ {\rm ker}D_{10}$, we have
\begin{align}
&\partial W_1 +2i [L_-, W_2]+\frac{s}{c}(W_1+2[L_3,W_1])=0, \nonumber\\
&\partial W_2 -2i [L_+, W_1]+\frac{s}{c}(W_2-2[L_3,W_2])=0,
\label{ker hyper}
\end{align}
where $s=\sinh \tau$ and $c=\cosh \tau$.
To analyze the structure of these equation, we use the 
fuzzy spherical harmonics, which behave nicely under the adjoint action 
of $L_i$. See \cite{Ishiki:2006yr,Ishii:2008tm,Ishii:2008ib} for 
the definition.
For the vacuum of the form (\ref{irrdec})
we can decompose $W_i (i=1,2)$ to the block components 
$\{W_i^{(s,t)} |s,t=1,2, \cdots, \Lambda \} $.
We then expand each block with the fuzzy spherical harmonics 
$\hat{Y}_{Jm(j_s,j_t)}$ as
\begin{align}
W_i^{(s,t)}=\sum_{J=|j_s-j_t|}^{j_s+j_t}\sum_{m=-J}^{J}
W_{iJm}^{(s,t)}\otimes \hat{Y}_{Jm(j_s,j_t)}.  \;\;\; (i=1,2)
\label{harmonic expansion for W}
\end{align}
Then, (\ref{ker hyper}) becomes
\begin{align}
&\partial W_{1Jm}^{(s,t)}+\frac{s}{c}(1+2m)W_{1Jm}^{(s,t)}
+2i\delta_- W_{2Jm+1}^{(s,t)}=0, \nonumber\\
&\partial W_{2Jm}^{(s,t)}+\frac{s}{c}(1-2m)W_{2Jm}^{(s,t)}
-2i\delta_+ W_{1Jm-1}^{(s,t)}=0,
\end{align}
where $\delta_{\pm}= \sqrt{(J\pm m)(J\mp m+1)}$.
It is easy to check that (\ref{hantei}) 
is satisfied only by $W_{1JJ}^{(s,t)}$ and $W_{2J-J}^{(s,t)}$. 
Indeed, these modes have eigenvalues $(2J+1) \tanh \tau$ which 
satisfy (\ref{hantei}).
Thus, only $W_{1JJ}^{(s,t)}$ and $W_{2J-J}^{(s,t)}$ 
and their complex conjugates contribute to the index. 

%The equations for the other modes can be rewritten in the form of 
%(\ref{df=0}), where $f=(W_{1Jm}^{(s,t)},W_{2Jm+1}^{(s,t)})^T$ and 
%\begin{align}
%A=\left(
%\begin{array}{cc}
%\frac{s}{c}(2m+1) & 2i\delta_-   \\
%-2i\delta_- & -\frac{s}{c}(2m+1)    \\
%\end{array}
%\right)
%\label{A hyper}
%\end{align}
%for $m=-J,-J+1,\cdots,J-1$.
%The eigenvalues of (\ref{A hyper}) do not satisfy (\ref{hantei}).
%Thus, we find that only $W_{1JJ}^{(s,t)}$ and $W_{2J-J}^{(s,t)}$ 
%and their complex conjugates contribute to the index. 

Then, let us consider the contribution from fermions, 
$\{\Upsilon_i, i=1,2,3,4\}$. 
We introduce complex fields as 
\begin{align}
\xi_{1}=\Upsilon_1+i\Upsilon_4, \;\;\; \xi_2=\Upsilon_3+i\Upsilon_2,
\end{align}
and expand their block components by the spherical harmonics 
as we did above. Then, we can obtain 
\begin{align}
&\partial \xi_{1Jm}^{(s,t)}+\frac{2sm}{c}\xi_{1Jm}^{(s,t)}+2\delta_+
\xi_{2Jm-1}^{(s,t)}=0, \nonumber\\
&\partial \xi_{2Jm}^{(s,t)}-\frac{2sm}{c}\xi_{2Jm}^{(s,t)}+2\delta_-
\xi_{1Jm+1}^{(s,t)}=0,
\end{align}
for $\xi_{1},\xi_2 \in {\rm coker}D_{10} $. In this case, there is no 
eigenvalue satisfying (\ref{hantei}).
Hence, these modes have no contribution to the index.

Thus, we find that only $W_{1JJ}^{(s,t)}$ and $W_{2J-J}^{(s,t)}$  
and their complex conjugates contribute to the index. 
The eigenvalues of $R$ for these modes are
$r= 2(\pm (2J+1) + i(q_{si}-q_{tj}))$, and thus we obtain (\ref{z hyper}).

%The eigenvalues of $R$ for these fields are read off from
%\begin{align}
%RW^{(s,t)}_{1}&=2W_{1}^{(s,t)}+[\hat{\phi},W_{1} ]^{(s,t)}
%\nonumber\\
%&= \sum_{J,m}2\left\{ (1+2m)W^{(s,t)}_{1Jm}+i(M_sW^{(s,t)}_{1Jm}-
%W^{(s,t)}_{1Jm}M_t) \right\}
%\otimes \hat{Y}_{Jm(j_s,j_t)}
%\end{align}
%and so on. 
%Thus, we find that the contribution to (\ref{z pert}) 
%from the hypermultiplet is given, up to an overall constant, by 
%\begin{align}
%\prod_{s,t=-\Lambda/2}^{\Lambda/2}\prod_{J=|j_s-j_t|}^{j_s+j_t}
%\prod_{i=1}^{N_s}\prod_{j=1}^{N_t}\frac{1}{(2J+1)^2+(m_{si}-m_{tj})^2}.
%\label{z hyper}
%\end{align}

\subsubsection*{Vector multiplet}
Next, we consider the vector multiplet.
We first calculate ${\rm dim }({\rm ker}D_{10})$.
For $\{Y_A,\tilde{a}_0,b_0 | A=1,2,3,4,9\} \in {\rm ker}D_{10}$, 
we have 
\begin{align}
&F+b_0=0, \label{g1}  \\
&\tilde{a}_0+2\left[\hat{Y}_{A},\frac{1}{\epsilon\epsilon}
[\hat{Y}_{A},v^4Y_4+v^9Y_9   ]\right]
+\left[
\left[
\hat{Y}_{A},\frac{1}{\epsilon \epsilon}Y_{A}
\right], -2iM+v^4\hat{Y}_4 \right]=0, \label{g2} \\
&c(2Y_4-i[\hat{Y}_2,Y_3]+i[\hat{Y}_3,Y_2])-s(\partial Y_4 +i[\hat{Y}_4,Y_1])
-\partial Y_9 =0, \label{g3} \\
& c(\partial Y_3 +i[\hat{Y}_3,Y_1])-s(2Y_3+i[\hat{Y}_2,Y_4]-i[\hat{Y}_4,Y_2])
-i[\hat{Y}_2,Y_9]=0, \label{g4} \\
& c(\partial Y_2 +i[\hat{Y}_2,Y_1])-s(2Y_2-i[\hat{Y}_3,Y_4]+i[\hat{Y}_4,Y_3])
+i[\hat{Y}_3,Y_9]=0. \label{g5}
\end{align}

To simplify the equations, 
let us consider the limit $\tau \rightarrow \pm \infty$ in (\ref{g1}).
Since $F \rightarrow 0$ in this limit, we obtain $b_0=0$. 
Noticing that $b_0$ has only the constant mode, 
by using (\ref{g1}) again, we find that $F$ should be vanishing 
for arbitrary point on $R$, namely,
\begin{align}
F=\sum_{a=1}^4\left[\hat{Y}_a, \frac{1}{\cosh \tau}Y_a \right]=0.
\label{F=0}
\end{align}
Similarly, $\tilde{a}_0=0$ follows from (\ref{g2}).  
By substituting these vanishing conditions to (\ref{g2}), 
we obtain,
\begin{align}
-\partial \left(
\frac{1}{c}\partial (Y_4-sY_9) \right)
+\frac{4}{c}\sum_{i=1}^3[L_{i},[L_{i},Y_4-sY_9]]=0.
\label{X4-sX9}
\end{align}
This equation implies $Y_4-sY_9=0$ as follows.
Putting $f = Y_4-sY_9$, the equation (\ref{X4-sX9}) 
has the form $\partial^2 f -\frac{s}{c}\partial f-4J(J+1)f=0$, where 
$J(J+1)$ is the eigenvalue of $[L_{i},[L_{i}, \;\; ]]$. 
From the boundary condition, $f/c$ should vanish at infinity. 
Then, it follows that
\begin{align}
0=\int d\tau \partial \left( \frac{1}{c^2}f\partial f \right)
=\int dx 
\left[ 
\left( \frac{\partial f }{c} \right)^2
+\left(\frac{4J(J+1)-1}{c^2}+\frac{3}{2c^4} \right)f^2
\right].
\end{align}
For $J\neq 0$, the right-hand side is a sum of positive definite terms
and hence $f$ itself must be zero. 
For $J=0$, the equation (\ref{X4-sX9}) is just 
$\partial ((\partial f)/c)=0$. By integrating this equation
under the boundary condition $f/c \rightarrow 0$, 
we find that $f$ is constant.
We then consider (\ref{g3}) with $f$ constant.  
From this equation, we can easily obtain $Y_4=Y_9=0$ for $J=0$. 
Therefore, the relation $Y_4 = sY_9$ holds for any $J$.

Then, by eliminating $Y_4$ by $Y_4 = sY_9$,
the equations (\ref{g1}), (\ref{g3}), (\ref{g4}), (\ref{g5}) become
\begin{align}
&-i\partial Y_1 +i \frac{s}{c}Y_1+[L_+,Y_-]+[L_-,Y_+]+2s[L_3,Y_9]=0, 
\nonumber\\
&-[L_+,Y_-]+[L_-,Y_+]+sY_9-c\partial Y_9+2i\frac{s}{c}[L_3,Y_1]=0, \nonumber\\
&c(\partial Y_+ -2i[L_+,Y_1])-s(2Y_+ -2[L_3,Y_+])-2c^2[L_+,Y_9]=0, \nonumber\\
&c(\partial Y_- -2i[L_-,Y_1])-s(2Y_- +2[L_3,Y_-])+2c^2[L_-,Y_9]=0,
\label{g1345}
\end{align}
where $Y_{\pm}=Y_2 \pm i Y_3$.

We then make a redefinition $Y_9'=c Y_9$\footnote{ 
Note that $Y_9'$ does not necessarily vanish at infinities.
However, only when $Y_9'$ vanishes at infinities, 
(\ref{g1345}) has nontrivial solutions.
So we assume that 
$Y_9' \rightarrow 0 $ as $\tau \rightarrow \pm \infty$.},
and expand each block component of $Y_{\pm}, Y_1, Y_9'$ 
by the fuzzy spherical harmonics.
For $f=(Y^{+(s,t)}_{Jm+1}/\sqrt{2},
Y^{-(s,t)}_{Jm-1}/\sqrt{2},iY^{1(s,t)}_{Jm},Y^{'9(s,t)}_{Jm})^T$
$(m=-J+1,-J+2,\cdots,J-1, \ J\geq 1)$,
the equations (\ref{g1345}) take the same form as 
(\ref{df=0}), where $A$ is given by
\begin{align}
A=
\left(
\begin{array}{cccc}
\frac{2ms}{c} & 0 & -\sqrt{2}\delta_- & -\sqrt{2}\delta_- \\
0 & -\frac{2ms}{c} & -\sqrt{2}\delta_+ & \sqrt{2}\delta_+ \\
-\sqrt{2}\delta_- & -\sqrt{2}\delta_+ & -\frac{s}{c} &-\frac{2ms}{c} \\
-\sqrt{2}\delta_- & \sqrt{2}\delta_+ & -\frac{2ms}{c} & -\frac{2s}{c} \\
\end{array}
\right).
\end{align} 
This matrix does not have any eigenvalues, which satisfy (\ref{hantei}).
Hence, we find that the bosonic fields in the vector multiplet 
do not contribute to the index.

Let us apply the same analysis to the fermions. 
For 
$(C,\tilde{C},\Upsilon_5,\Upsilon_6,\Upsilon_7) \in {\rm coker}D_{10}$,
we have
\begin{align}
&-\frac{1}{c}\partial \tilde{C}+\frac{1}{c}[iM-2L_3,\partial C]
-8 s [L_3, \Upsilon_5]-8c[L_2,\Upsilon_6]+8c[L_1,\Upsilon_7]=0,  
\nonumber\\
&\frac{1}{c}[L_1,\tilde{C}]-\frac{1}{c}[L_1,[iM-2L_4,C]]
+4ic[L_2,\Upsilon_5]-4is[L_3,\Upsilon_6]-2c\partial \Upsilon_7 -6s\Upsilon_7 =0,\nonumber\\
&\frac{1}{c}[L_2,\tilde{C}]-\frac{1}{c}[L_2,[iM-2L_4,C]]
-4ic[L_1,\Upsilon_5]-4is[L_3,\Upsilon_7]+2c\partial \Upsilon_6 +6s\Upsilon_6 =0,\nonumber\\
&\frac{1}{c}[L_3,\tilde{C}]+\partial \left(\frac{1}{c}\partial C \right) 
-\frac{4}{c}\sum_{i=1}^3[L_{i},[L_{i},C]]-\frac{1}{c}[L_3,[iM-2L_3,C]] 
+2s \partial \Upsilon_5 +6c \Upsilon_5 \nonumber\\
& +4is[L_1,\Upsilon_6]+4is[L_2,\Upsilon_7] =0, 
\nonumber\\
&-s \partial \left(\frac{1}{c}\partial C \right) 
+\frac{4s}{c}\sum_{i=1}^3[L_{i},[L_{i},C]]+2\partial \Upsilon_5
+4i[L_1,\Upsilon_6]+4i[L_2,\Upsilon_7]=0.
\label{coker gauge}
\end{align}
%From the coefficients of $\tilde{a}_0$ and $b_0$
%in (\ref{linearized action}), we have 
%\begin{align}
%\int_{-\infty}^{\infty}d\tau C^{(s,s)}_{00}(\tau)=
%\int_{-\infty}^{\infty}d\tau \tilde{C}^{(s,s)}_{00}(\tau)=0,
%\label{0 mode is 0}
%\end{align}
%where $C^{(s,s)}_{00}, \tilde{C}^{(s,s)}_{00}$ 
%the zero mode of the fuzzy sphere which 
%exists only in the diagonal blocks.
We make some redefinitions as  
$\tilde{C}'=(\tilde{C}-[iM-2L_4,C])/(2\sqrt{2} c)$,
$C'=C/c$, $\Upsilon_5'=\sqrt{2}\Upsilon_5$
and also introduce complex fields, 
$\Upsilon_{\pm}=\Upsilon_6 \pm i\Upsilon_7$.
%Note that these fields also vanish at infinity.
With this notation, we can write (\ref{coker gauge}) as
\begin{align}
&\partial C' -d =0,
\nonumber\\
&\partial d +\frac{3s}{c}d +2C'-4\sum_{i=1}^3[L_{i},[L_{i},C']]
+\frac{2\sqrt{2}}{c^2}[L_3,\tilde{C}'] 
+\frac{3\sqrt{2}}{c^2}\Upsilon'_5 =0,
\nonumber\\
&\partial \Upsilon_+ -\sqrt{2}i[L_+,\tilde{C}']
-\sqrt{2}i[L_+,\Upsilon_5']+\frac{3s}{c}\Upsilon_+ 
-\frac{2s}{c}[L_3,\Upsilon_+]=0,
\nonumber\\
&\partial \Upsilon_- +\sqrt{2}i[L_-,\tilde{C}']
-\sqrt{2}i[L_-,\Upsilon_5']+\frac{3s}{c}\Upsilon_- 
+\frac{2s}{c}[L_3,\Upsilon_-]=0,
\nonumber\\
&\partial \tilde{C}' +\frac{2s}{c}\tilde{C}' 
+\frac{2s}{c}[L_3,\Upsilon_5']-\sqrt{2}i([L_+,\Upsilon_-]-[L_-,\Upsilon_+])=0,
\nonumber\\
&\partial \Upsilon_5' +\frac{2s}{c}[L_3,\tilde{C}']
+\frac{3s}{c}\Upsilon_5'
+\sqrt{2}i([L_+,\Upsilon_-]+[L_-,\Upsilon_+])=0,
\end{align}
where a new field $d$ is introduced to make the equations first order.

We then expand each block component by fuzzy spherical harmonics.
For $f=(C'{}^{(s,t)}_{Jm},d^{(s,t)}_{Jm},
\Upsilon^{+(s,t)}_{Jm+1},
\Upsilon^{-(s,t)}_{Jm-1},
\Upsilon'{}^{5(s,t)}_{Jm},
\tilde{C}'{}^{(s,t)}_{Jm})^T$  
$(m=-J+1,-J+2,\cdots,J-1)$, 
the above equation can be written in the form of 
(\ref{df=0}), where
\begin{align}
A=
\left( 
\begin{array}{cccccc}
0 & -1 & 0 & 0 & 0 & 0 \\
\frac{3s}{c} & 2-4J(J+1) & 0 
& 0 & \frac{3\sqrt{2}}{c^2} & \frac{3\sqrt{2}m}{c^2}  \\
0 & 0 & \frac{s}{c}(1-2m) 
& 0 & -\sqrt{2}i \delta_- & -\sqrt{2}i \delta_- \\
0 & 0 & 0 
& \frac{s}{c}(1+2m) & -\sqrt{2}i\delta_+ & \sqrt{2}i\delta_+  \\
0 & 0 & \sqrt{2}i\delta_- 
& \sqrt{2}i\delta_+ & \frac{3s}{c} & \frac{2ms}{c} \\
0 & 0 & \sqrt{2}i\delta_- 
& -\sqrt{2}i\delta_+ & \frac{2ms}{c} & \frac{2s}{c} \\ 
\end{array}
\right).
\label{A for coker}
\end{align}
It is easy to see that there is no eigenvalues of $A$ that satisfy (\ref{hantei}). Hence, these modes have no contribution to the index.

On the other hand, the highest momentum modes
$f=(C'{}^{(s,t)}_{JJ},d^{(s,t)}_{JJ}, \Upsilon^{-(s,t)}_{JJ-1},
\Upsilon'{}^{5(s,t)}_{JJ},\tilde{C}'{}^{(s,t)}_{JJ})^T$ 
have a nontrivial contribution. They 
satisfy (\ref{df=0}) where $A$ is given by a $5\times 5$ matrix
obtained by eliminating the fifth row and column (namely, those 
for $\Upsilon_+$) and putting $m=J$ in (\ref{A for coker}).
Then, we can find 
that there is just one eigenvalue which satisfies (\ref{hantei}).
%Therefore, together with the complex conjugate,
%$f=(C'{}^{(s,t)}_{J-J},d^{(s,t)}_{J-J}, \Upsilon^{+(s,t)}_{J-J+1}
%, \Upsilon'{}^{5(s,t)}_{J-J}, \tilde{C}'{}^{(s,t)}_{J-J})^T$, 
In the same way, we can see that the modes with $m=-J$ have the same 
structure\footnote{In fact, these modes are the complex conjugate of the 
highest modes with $m=J$.}. The eigenvalues of $R$ for 
these modes are $r= 2(\pm 2J + i(q_{si}-q_{tj}))$.
This contribution gives the first line of (\ref{z vector}).

Finally, $\Upsilon^{+(s,t)}_{J-J}$ and 
$\Upsilon^{-(s,t)}_{JJ}$ satisfy the closed equation 
$\partial \Upsilon +\frac{(2J+3)s}{c}\Upsilon =0 $. 
Then (\ref{hantei}) is satisfied and hence they contribute to the index. 
The eigenvalues of $R$ are $r =2(\pm (2J+2) + i(q_{si}-q_{tj}))$.
This gives the second line of (\ref{z vector}).

%%%%%%%%%%%%%%%%%%%%%%%%%%%%%%%%%%%%%%%%%%%%%%%%%%%%%%%%%%%%%%%
\section{The saddle point equation}
\label{The saddle point equation}
In this appendix, we derive (\ref{spe}).
We start with the effective action of (\ref{one matrix}), 
\begin{align}
S_{eff}= &\beta \left(1-\int_{-q_m}^{q_m} dq \rho (q) \right)
+\frac{2N_5}{\lambda} \int_{-q_m}^{q_m} dq q^2 \rho (q)  
\nonumber\\
&-\frac{1}{2}\int_{-q_m}^{q_m} dq \int_{-q_m}^{q_m} dq' 
\rho(q) \rho (q') \sum_{J=0}^{N_5-1}
\log \frac{\{(2J+2)^2 +(q-q')^2 \}\{(2J)^2 +(q-q')^2 \} }
{\{(2J+1)^2 +(q-q')^2 \}^2}.
\label{effective action}
\end{align}
Here, $\lambda=g^2 N_2$ is the 't Hooft coupling and 
$\beta$ is the Lagrange multiplier for the normalization 
of the eigenvalue density (\ref{normalization of rho}).
The saddle point equation is obtained by differentiating 
$S_{eff}$ with respect to $\rho(q)$ and is given by
\begin{align}
\beta = \frac{2N_5}{\lambda}q^2  
- \int^{q_m}_{-q_m} dq' \rho (q') \sum_{J=0}^{N_5-1}
\log \frac{\{(2J+2)^2 +(q-q')^2 \}\{(2J)^2 +(q-q')^2 \} }
{\{(2J+1)^2 +(q-q')^2 \}^2}.
\label{spe 1}
\end{align}
Here the integral of $q'$ should be understood as the principal value.

We first consider the following identity,
\begin{align}
\log \tanh^2 \left( \frac{\pi x}{2} \right)
= \log \left( \frac{\pi x}{2} \right)^2 
+ \sum_{J=1}^{\infty} \left(1+ \frac{x^2}{(2J)^2} \right)^2
- \sum_{J=1}^{\infty} \left(1+ \frac{x^2}{(2J-1)^2} \right)^2,
\end{align}
which follows from the infinite product expression of 
the hyperbolic sine and cosine functions.
By using this identity, we find that the second term in (\ref{spe 1})
can be written as 
\begin{align}
-\int^{q_m}_{-q_m} dq' \rho (q' )
\left[ 
\log \tanh^2 \left( \frac{\pi (q-q')}{2} \right)
-\sum_{J=N_5}^{\infty}
\log \frac{\{(2J+2)^2 +(q-q')^2 \}\{(2J)^2 +(q-q')^2 \} }
{\{(2J+1)^2 +(q-q')^2 \}^2}
\right]
\label{spe 2}
\end{align}
up to a constant term. We ignore the constant term since 
it can always be absorbed by a redefinition of $\beta$.
In the regime where $N_5$ is finite but 
$\lambda$ is very large, $q_m$ also becomes very large.
To see the $q_m$-dependence clearly, 
let us rescale the variables as $q=q_m \xi$.
From the fact that 
\begin{align}
q_m \log \tanh^2 \frac{\pi q_m \xi }{2} \rightarrow -\pi \delta(\xi) \;\; 
(q_m \rightarrow \infty),
\label{logtanh and delta}
\end{align}
we find that the first term in (\ref{spe 2}) is equal to $\pi \rho(q)$
in this limit.
In the second term in (\ref{spe 2}), we approximate the discrete sum with 
a continuous integral by replacing $J/q_m \rightarrow \eta$ and
$\sum_{J=N_5}^{\infty} \rightarrow q_m \int_{N_5/q_m}^\infty d\eta$.
Then, the second term can be evaluated as
\begin{align}
&- 2 \int^{1}_{-1} d\xi' \rho(q_m \xi') 
\int_{N_5/q_m}^\infty d\eta 
\left[
\frac{4\eta^2-(\xi-\xi')^2}{(4\eta^2 +(\xi-\xi')^2)^2}
+{\cal O}(1/q_m)
\right]
\nonumber\\
&\simeq 
- \int^{q_m}_{-q_m}dq' \rho (q') \frac{2N_5}{(2N_5)^2+(q-q')^2} .
\end{align}
Thus, in the strongly coupled regime, the saddle point equation (\ref{spe 1})
is reduced to (\ref{spe}).

%%%%%%%%%%%%%%%%%%%%%%%%%%%%%%%%%%%%%%%%%%%%%%%%%%%%%%%%%%%%%%%%%%%%%%%%%%%%%%%
\section{Solving the saddle point equation}
In this appendix, we construct solutions of the saddle point equations of 
the eigenvalue integrals obtained by the localization.
\subsection{For the simplest partition}
\label{Solving the saddle point equation}
%%%%%%%%%%%%%%%%%%%%%%%%%%%%%%%%%%%%%%%%%%%%%%%%%%%%%%%%%%%%%%%%%%%%%%%%%%%%%%%
Here, we derive (\ref{density solution}).
We first rewrite (\ref{spe}) into a more tractable form.
We define the resolvent by
\begin{align}
\omega(z)= \int_{-q_m}^{q_m} dq \frac{\rho (q)}{z-q}.
\end{align}
For $q\in [-q_m, q_m]$, this satisfies 
\begin{align}
\omega(q \pm i0) = P \int^{q_m}_{-q_m}dq' \frac{\rho (q')}{q-q'} 
\mp \pi i \rho (q),
\label{omega pm}
\end{align}
where $P \int^{q_m}_{-q_m}$ denotes the principal value. 
Note that the last term in (\ref{spe}) can be written as
$\frac{1}{2i}\left\{ 
\omega(q-2iN_5) -\omega(q+2iN_5)
\right\}$.
By using this and (\ref{omega pm}), we rewrite (\ref{spe}) as
\begin{align}
\beta = \frac{1}{2i}
\left\{ 
\omega(q+2iN_5)-\omega(q+i0)
\right\}
-
\frac{1}{2i}
\left\{ 
\omega(q-2iN_5)-\omega(q-i0)
\right\}
+\frac{2N_5}{\lambda} q^2.
\label{spe with omega}
\end{align}
When $q_m$ is large compared to $N_5$, we can expand $\omega(q \pm 2iN_5)$ 
as
\begin{align}
\omega(q\pm 2iN_5) = \omega(q \pm i0) \pm 2iN_5 \omega'(q\pm i0) + \cdots.
\end{align}
The convergence of this expansion can be seen clearly 
if one rescales the variable as $q= q_m \xi$, as we did in 
appendix \ref{The saddle point equation}.
Thus, in the large-$q_m$ limit, 
the equation (\ref{spe with omega}) becomes 
\begin{align}
\beta = N_5 
\left\{
\omega'(q+i0) + \omega' (q-i0)
\right\}
+\frac{2N_5}{\lambda} q^2.
\label{spe in M5 limit}
\end{align}
By integrating this equation, we obtain
\begin{align}
\omega(q+i0)+ \omega(q-i0)= \frac{\beta }{N_5}q -\frac{2}{3\lambda }q^3,
\label{spe with omega 2}
\end{align}
where we have set the integration constant to be zero because of the symmetry 
under $q \rightarrow -q$.

The equation (\ref{spe with omega 2}) is identical with the equation of motion
of the quartic one matrix model.
Hence, the solution takes the same form as the quartic matrix model, where 
the resolvent is written as
\begin{align}
\omega(z) = \frac{1}{2}\left\{\omega(q+i0)+ \omega(q-i0) \right \}
+ (a+b z^2) \sqrt{z^2- q_m^2}.
\label{ansatz}
\end{align}
We substitute (\ref{spe with omega 2}) into this expression.
Then, the asymptotic behavior of the resolvent, 
$\omega (z) \rightarrow \frac{1}{z}\; (z\rightarrow \infty)$, 
gives three conditions, which enable us to express 
$a$, $b$ and $\beta$ in terms of $q_m$:
\begin{align}
a=-\frac{2}{q_m^2}+ \frac{2q_m^2}{3 \lambda}, \;\;\;\;\;
b=\frac{1}{3\lambda}, \;\;\;\;\;
\frac{\beta}{N_5}= \frac{4}{q_m^2}+\frac{q_m^2}{2 \lambda}.
\end{align}
Thus, the resolvent is finally determined as
\begin{align}
\omega (z)= 
\left(
\frac{2}{q_m^2}
+ \frac{q_m^2}{4\lambda}
\right)z 
-\frac{1}{3\lambda} z^3
-
\left(
\frac{2}{q_m^2}
+\frac{q_m^2}{12\lambda}
-\frac{z^2}{3\lambda }
\right)\sqrt{z^2-q_m^2}.
\label{resolvent solution}
\end{align}
The eigenvalue density is given by the discontinuity of
\eqref{resolvent solution} as
\begin{align}
\rho (q)= \frac{1}{\pi} \left(
\frac{2}{q_m^2}
+\frac{q_m^2}{12\lambda}
-\frac{q^2}{3\lambda }
\right)
\sqrt{q_m^2-q^2}.
\label{solution for eigenvalue density}
\end{align}
Note that in order for $\rho (q)$ to be positive for any $q \in [-q_m, q_m]$, 
$q_m$ has to satisfy 
\begin{align}
q_m^4 \leq 8\lambda.
\label{region of qm}
\end{align}

Finally, we determine the value of $q_m$ from the action principle.
By using the saddle point equation, we can reduce 
the effective action (\ref{effective action}) to 
\begin{align}
S_{eff}/(N_2)^2= \frac{N_5}{\lambda } \int_{-q_m}^{q_m} q^2 \rho (q)
+ \frac{\beta }{2}.
\end{align}
By evaluating this using 
(\ref{solution for eigenvalue density}),
we obtain the on-shell value of the effective action as 
\begin{align}
S_{eff}/(N_2)^2= \frac{2N_5}{q_m^2}
\left(
1+\frac{q_m^4}{4\lambda}
-\frac{q_m^8}{192\lambda^2}
\right).
\label{onshell action}
\end{align}
In the region (\ref{region of qm}), the minimum of $S_{eff}$ 
is realized at 
\begin{align}
q_m^4 = 8 \lambda.
\end{align}
By substitute this into \eqref{solution for eigenvalue density}, 
we obtain (\ref{density solution}).

\subsection{For the generic partition}
\label{Solving the saddle point equation 2}

Here, we construct a solution to
(\ref{saddle for general partition}) in the decoupling limit of M5-branes.

In the decoupling limit, by applying the same computation 
that we used to derive 
(\ref{spe in M5 limit}), we can reduce the saddle point equations
(\ref{saddle for general partition}) to
\begin{align}
\frac{1}{2}\sum_{t=1}^{\Lambda}
\left(
n_s+n_t-|n_s-n_t|
\right)
(\omega_t'(q+i0)+\omega_t'(q-i0)) = \beta_s -\frac{2n_s}{g^2}q^2,
\label{generic 1}
\end{align}
where we have defined the resolvent as 
\begin{align}
\omega_s(z) = \int^{q_s}_{-q_s}dq \frac{\rho_s(q)}{z-q}.
\end{align}

Without loss of generality, we assume that $n_s$ 
in the decomposition (\ref{irrdec}) are ordered as $n_1>n_2> \cdots > 
n_\Lambda$. We also assume that 
\begin{align}
q_\Lambda>q_{\Lambda -1}> \cdots > q_1.
\label{order of x}
\end{align}
Then, let us first consider the 
equation (\ref{generic 1}) with $s=\Lambda$,
\begin{align}
n_\Lambda \sum_{t=1}^\Lambda (\omega_t'(q+i0)+\omega_t'(q-i0))
= \beta_\Lambda -\frac{2n_\Lambda}{g^2}q^2 ,  \;\;\; q \in [-q_\Lambda, q_\Lambda ].
\label{generic 2}
\end{align}
Under the assumption (\ref{order of x}), it makes sense to consider 
\begin{align}
\hat{\rho}_{\Lambda}(q) = \sum_{s=1}^{\Lambda} \rho_s(q),
\end{align}
which has the support $[-q_\Lambda, q_\Lambda ]$ and is normalized as 
\begin{align}
\int^{q_\Lambda}_{-q_\Lambda} dq \hat{\rho}_{\Lambda}(q) = \sum_{s=1}^{\Lambda}
N_2^{(s)}.
\end{align}
In terms of $\hat{\rho}_\Lambda (q)$, (\ref{generic 2}) can be simply 
written as
\begin{align}
n_\Lambda (\hat{\omega}_\Lambda'(q+i0)+\hat{\omega}_\Lambda'(q-i0))
= \beta_\Lambda -\frac{2n_\Lambda}{g^2}q^2 ,  \;\;\; q \in [-q_\Lambda, q_\Lambda ],
\label{generic 3}
\end{align}
where $\hat{\omega}_\Lambda (z)$ is the resolvent for 
$\hat{\rho}_{\Lambda}(q)$.
Since (\ref{generic 3}) takes the same form as (\ref{spe in M5 limit}), 
the solution for $\rho_\Lambda(q)$ is also given by the same form as 
(\ref{solution for eigenvalue density}). We will determine 
$q_\Lambda$ below.
%\begin{align}
%&\hat{\rho}_\Lambda(q) = \frac{8^{3/4}\sum_{s=1}^{\Lambda}N_2^{(s)}}
%{3\pi \lambda_\Lambda^{1/4}}\left[ 
%1-\frac{q^2}{q^2_\Lambda} 
%\right]^{\frac{3}{2}}, \;\;\;
%q_\Lambda = (8\lambda_\Lambda)^{1/4}, \;\;\; \lambda_\Lambda := g^2
%\sum_{s=1}^{\Lambda}N_2^{(s)}.
%\end{align}

Next, we solve (\ref{generic 1}) with $s< \Lambda$. 
Let us consider the difference between (\ref{generic 1}) with 
$s=r$ and (\ref{generic 1}) with $s=r+1$ on the support $q \in [-q_r, q_r]$, 
where $r \in \{1,2,\cdots, \Lambda-1\} $.
This leads to 
\begin{align}
(n_r-n_{r+1})\sum_{t=1}^r (\omega_t'(q+i0)+\omega_t'(q-i0))
= \beta_r -\beta_{r+1} -\frac{2(n_r-n_{r+1})}{g^2} q^2, 
\;\;\; q \in [-q_r, q_r ].
\label{generic 4}
\end{align} 
Note that $\rho_s(q)$ with $s>r$ does not appear in this equation. 
We introduce new variables 
\begin{align}
\hat{\rho}_r (q) = \sum_{s=1}^r \rho_s(q), 
\end{align}
which are normalized as 
\begin{align}
\int^{q_r}_{-q_r} dq \hat{\rho}_{r}(q) = \sum_{s=1}^{r}N_2^{(s)}.
\end{align}
In terms of $\hat{\rho}_r(q)$, (\ref{generic 4}) becomes 
\begin{align}
(n_r-n_{r+1})(\hat{\omega}_r'(q+i0)+\hat{\omega}_r'(q-i0))
= \beta_r -\beta_{r+1} -\frac{2(n_r-n_{r+1})}{g^2} q^2, 
\;\;\; q \in [-q_r, q_r ].
\end{align}
Again, this is the same form as (\ref{spe in M5 limit}), so that 
the solution for $\hat{\rho}_r$ is given by the same form as
(\ref{solution for eigenvalue density}).

Finally we determine $q_r$. By using the equation of motion for 
$\rho_s$, the on-shell can be computed as
\begin{align}
S_{eff}= \sum_{s=1}^\Lambda \frac{2(n_s-n_{s+1})}{q_s^2}
\left(
1+\frac{q_s^4}{4\lambda_s}
-\frac{q_s^8}{192\lambda_s^2}
\right),
\label{onshell action general}
\end{align}
where $\lambda_s= g^2 \sum_{r=1}^s N_2^{(r)}$. 
Thus, the minimum is given by $q_s = (8\lambda_s)^{1/4}$. 
Thus, we obtained (\ref{solution for generic partition}).

%\begin{align}
%&\hat{\rho}_r(q) = \frac{8^{3/4}\sum_{s=1}^{r}N_2^{(s)}}
%{3\pi \lambda_r^{1/4}}\left[ 
%1-\frac{q^2}{q^2_r} 
%\right]^{\frac{3}{2}}, \;\;\;
%q_r = (8\lambda_r)^{1/4}, \;\;\; \lambda_r := g^2
%\sum_{s=1}^{r}N_2^{(s)}.
%\end{align}
%Thus, we obtained (\ref{solution for generic partition}).

\section{Eigenvalue distribution in the M2-brane limit}
\label{Eigenvalue distribution in the M2-brane limit}
In this appendix, we solve the 
eigenvalue integral (\ref{D2 partition function}) for large $N_2$.
Putting $a= g^2 N_2/N_5$, 
we consider the scaling limit such that 
$N_5, N_2, a \rightarrow \infty$, 
$N_5/N_2 \rightarrow \infty $ and  $N_2/a \rightarrow \infty $.
We again assume that the typical value of eigenvalues is 
very large in this limit, 
since the Gaussian attractive force becomes weak. 

We introduce the eigenvalue density $\rho (q)$ as 
(\ref{eigenvalue distribution def}).
The effective action of \eqref{D2 partition function}
is written in terms of $\rho (q)$ as
\begin{align}
S_{eff}/(N_2)^2 = &\frac{2N_5}{\lambda} \int^{q_m}_{-q_m} 
dq q^2 \rho (q) - \frac{1}{2} \int^{q_m}_{-q_m} dq \int^{q_m}_{-q_m} dq'
\rho(q) \rho(q')
\log \tanh^2 \left( \frac{\pi (q -q')}{2} \right)
\nonumber\\
&+ \beta \left(1-
\int^{q_m}_{-q_m} dq \rho(q)
\right).
\end{align}
By applying (\ref{logtanh and delta}), we find that 
the action reduces to 
\begin{align}
S_{eff}/(N_2)^2 = \frac{2N_5}{\lambda} \int^{q_m}_{-q_m} 
dq q^2 \rho (q) + \frac{\pi}{2} \int^{q_m}_{-q_m} dq 
\rho(q)^2+ \beta \left(1-
\int^{q_m}_{-q_m} dq \rho(q)
\right).
\end{align}
The saddle point equation
is given by
\begin{align}
\beta = \pi \rho(q) + \frac{2N_5}{\lambda }q^2.
\label{spe D2}
\end{align}
Thus, $\rho(q)$ is a quadratic function in $q$ and $q_m$ is 
related to $\beta$ as 
\begin{align}
q_m^2 = \frac{\lambda}{2N_5}\beta.
\label{qm and mu D2}
\end{align}
From \eqref{normalization of rho},
\eqref{spe D2} and
\eqref{qm and mu D2},
we obtain 
\begin{align}
q_m = \left( 
\frac{3\pi \lambda}{8N_5}
\right)^{\frac{1}{3}}.
\end{align}
Thus, the typical value of the eigenvalues should be proportional to 
$\left( \frac{\lambda}{N_5} \right)^{1/3}$.
Note that 
this result is consistent with our assumption that 
the typical value of the eigenvalues 
is very large in the strong coupling region.

%%%%%%%%%%%%%%%%%%%%%%%%%%%%%%%%%%%%%%%%%%%%%%%%%%%%%%%%%%%%%%%%%%%%

\end{document}